\documentclass[preprint,prd,amsmath,amssymb]{revtex4}
\usepackage{graphicx}
\usepackage{subfigure}
\usepackage{dcolumn}

\newcommand{\be}{\begin{equation}}
\newcommand{\ee}{\end{equation}}

\newcommand{\Tr}{{\rm Tr}}
\newcommand{\bea}{\begin{eqnarray}}
\newcommand{\eea}{\end{eqnarray}}
\newcommand{\bml}{\begin{multline}}
\newcommand{\eml}{\end{multline}}

\newcommand{\LE}{{\cal E}}

\newcommand{\LH}{{\cal H}}

\newcommand{\nn}{\nonumber}
\newcommand{\bm}[1]{\mbox{\boldmath $#1$}}

\newcommand{\boldx}{\boldsymbol{x}}
\newcommand{\eqn}[1]{Eq.~(\ref{#1})}

\newcommand{\eqns}[2]{Eqs.~(\ref{#1}) and (\ref{#2})}

\newcommand{\fig}[1]{Fig.~\ref{#1}}
\newcommand{\figs}[2]{Figs.~\ref{#1} and~\ref{#2}}

\newcommand{\tabss}[2]{Tables~\ref{#1}---\ref{#2}}
\newcommand{\sect}[1]{Section~\ref{#1}}
\newcommand{\sects}[2]{Sections~\ref{#1} and~\ref{#2}}

\newcommand{\rcite}[1]{Ref.~\onlinecite{#1}}



\begin{document}

\title{SU($N$) Glueball Masses in 2+1 Dimensions}

\author{Jesse Carlsson}
\email{j.carlsson@physics.unimelb.edu.au}
\author{Bruce H.~J.~McKellar}
\email{b.mckellar@physics.unimelb.edu.au}
\affiliation{School of Physics, The University of Melbourne}
\date{\today}
\begin{abstract}
    We calculate the masses of the lowest lying eigenstates of
    improved SU(2), SU(3), SU(4) and SU(5) 
    Hamiltonian lattice gauge theory (LGT) in 2+1 dimensions
    using an analytic variational approach.  The ground state is
    approximated by a one plaquette trial state and mass gaps are
    calculated in the symmetric and antisymmetric sectors by
    minimising over a suitable basis of rectangular states. Analytic
    techniques are developed to handle the group integrals arising in
    the calculation.
\end{abstract}

\maketitle

\section{Introduction}

In this paper we calculate the lowest lying glueball masses for
SU($N$) LGT in 2+1 dimensions, with $N=2$, 3, 4 and 5, extending an
earlier paper which considered $N = 2$ and 3~\cite{Carlsson:2002ss}. We use 
Kogut-Susskind~\cite{Kogut:1975ag}, classically improved and tadpole improved Hamiltonians~\cite{Carlsson:2001wp} 
in their calculation and develop analytic techniques for the calculation of the required expectation values.

The outline of this paper is as follows. In \sect{suNintegrals} we
develop analytic techniques for use in the calculation of certain
group integrals for general SU($N$). 
When used as generating functions these integrals allow an analytic
treatment of the matrix elements appearing
in later sections. After defining our notation
in \sect{analyticSU(N)calcs} we fix the variational vacuum wave function
in \sect{fixingthevariational}. Following that we calculate
lattice specific heats in \sect{specheats-1} before studying 
SU($N$) glueball masses in \sect{massgaps}. \sect{concl}
contains our conclusions and a discussion of further work.

\section{Analytic techniques for SU($N$)}
\label{suNintegrals}
\subsection{The special cases of SU(2) and SU(3)}
\label{su3integrals}
In this section we recall two results which are useful in the
calculation of group integrals for the special cases of SU(2) and
SU(3). For further details we refer the reader to
\rcite{Carlsson:2002ss}. Before presenting these results the concept
of group integration needs to be introduced. For this purpose we  
introduce the one plaquette trial state, which we will use to simulate
the ground, or (perturbed) vacuum, state with energy $E_0$, 
\be
\begin{array}{c}\includegraphics{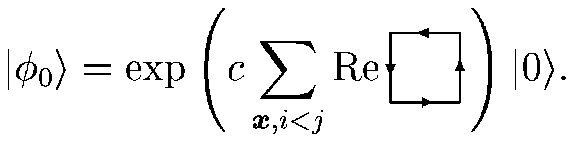}\end{array}
\label{oneplaquette}
\ee
Here, $|0\rangle $ is the strong coupling vacuum defined by
$\LE^\alpha_i(\bm{x})|0\rangle = 0$ for all $i$, $\bm{x}$ and $\alpha =
1,2,\ldots,N^2-1$. $\LE^\alpha_i(\bm{x})$ is the lattice
chromoelectric field on the directed link running 
from the lattice site labelled by $\bm{x}$ to the site labelled by $\bm{x}+a\bm{i}$. The directed square denotes the
traced ordered product of link 
operators, $U_i(\bm{x})$, around an elementary square, or plaquette, of the lattice,
\be
\begin{array}{c}\includegraphics{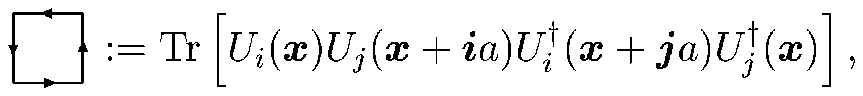}\end{array}
\ee
where $a$ is the lattice spacing.
With this notation understood, we can write the
expectation value of a plaquette as an SU($N$) group integral as follows
\be
\begin{array}{c}\includegraphics{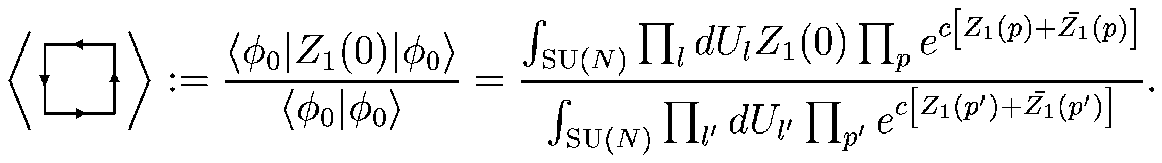}\end{array}
\label{groupintegralintro}
\ee
Here the products over $l$ and $l'$ extend over all links on the
lattice, while the products over $p$ and $p'$ extend over all
plaquettes on the lattice. In \eqn{groupintegralintro} we introduced
the notation $Z_1(p)$ to denote
the trace of plaquette $p$. For each integral in \eqn{groupintegralintro} the integration measure is 
given by the Haar measure (also called the invariant measure and less
commonly the Hurwitz measure)~\cite{Montvay:1994cy,Creutz:1984m}. 
For any compact group $G$, the Haar
measure is the unique measure $dU$ on $G$ which is left and right
invariant, 
\be
\int_G dU f(U) = \int_G dU f(V U) = \int_G dU f(U V) \hspace{1cm} \forall V
\in G
\label{LRinvariance}
\ee
and normalised,
\bea
\int_G dU =1. \label{normalisation}
\eea
In \eqn{LRinvariance} $f$ is an arbitrary function over $G$. 

In 2+1 dimensions the variables in \eqn{groupintegralintro} can be
changed from links to plaquettes with unit
Jacobian~\cite{Batrouni:1984rb}. The plaquettes then become
independent variables allowing the cancellation of all but one  
group integral in \eqn{groupintegralintro}. All that remains is
\be
\begin{array}{c}\includegraphics{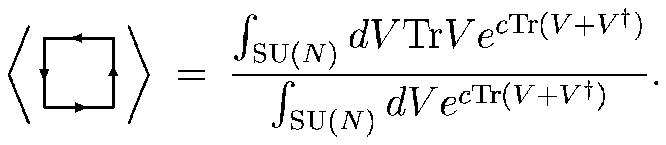}\end{array}
\ee
Here, $V$ is a plaquette variable with $\Tr V = Z_1(0)$.   
For the case of SU(2), analytic expressions for the 
plaquette expectation value in terms of modified Bessel 
functions have been used in variational 
calculations for almost 20 years. A key result~\cite{Arisue:1983tt} is
\bea
\int_{{\rm SU}(2)} dU e^{c \Tr U} = \frac{1}{c} I_1(2 c).
\label{su2gen}
\eea
Here $I_n$ is the $n$-th order modified Bessel function of the first
kind defined, for all integers $n$, by
\bea
I_n(2x) = \sum_{k=0}^\infty \frac{x^{2k+n}}{k!(k+n)!}.
\eea 

In an earlier paper, we showed that the corresponding SU(3) result 
follows simply from a paper of Eriksson, Svartholm and
Skagerstam~\cite{Eriksson:1981rq}, with the result:
\bea
\int_{{\rm SU}(3)}dU e^{c \Tr U + d \Tr U^\dagger} &=& 2 \sum_{k=0}^\infty \frac{1}{(k+1)!(k+2)!}
\sum_{l=0}^k\left(\!\begin{array}{c} 3k+3 \\ k-l \end{array}\!\right)
\frac{1}{l!}(cd)^{k-l}(c^3+d^3)^l . \label{Y}
\eea

In the next section we derive the general SU($N$) result as well as a
selection of other useful SU($N$) group integrals.

\subsection{The generalisation to SU($N$)}
\label{AnalyticresultsforSU}

\subsubsection{Introduction}

Much work has been carried out on the topic of integration over the classical
compact groups. The
subject has been studied in great depth in the context of random
matrices and combinatorics. Many analytic results in terms of determinants are
available for integrals of various functions over unitary, orthogonal 
and symplectic groups~\cite{Baik:2001}. Unfortunately similar results
for SU($N$) are not to our knowledge available. 
Their primary use has been in the study of Ulam's problem concerning
the distribution of the length of 
the longest increasing subsequence in permutation
groups~\cite{Rains:1998,Widom:2001}. Connections between random
permutations and Young tableaux~\cite{Regev:1981} allow an interesting
approach to combinatorial problems involving Young tableaux. A 
problem of particular interest is the counting of Young tableaux
of bounded height~\cite{Gessel:1990} which is closely related to the
problem of counting singlets in product representations mentioned in \sect{su3integrals}. 
Group integrals similar to those needed in
this paper have also appeared in studies of the distributions of the
eigenvalues of random matrices~\cite{Diaconis:1994,Widom:1999}. 

In the context of LGT not much has work been done in the last 20 years on the
subject of group integration. The last
significant development was due to Creutz who
developed a diagrammatic technique for calculating specific SU($N$)
integrals~\cite{Creutz:1978ub} using link variables. 
This technique allows strong coupling
matrix elements to be calculated for SU($N$)~\cite{ConradPhD}
and has more recently been used in the loop formulation of quantum
gravity where spin networks are of interest~\cite{DePietri:1997pj,Rovelli:1995ac,Ezawa:1997bv}.
  
In \sects{simpleintegral}{morecomplicated} 
we extend the results of \sect{su3integrals} to
calculate two important SU($N$) integrals. As generating functions
these integrals allow the evaluation of all expectation
values appearing in variational calculations of SU($N$) glueball masses in
2+1 dimensions. To calculate these generating functions we work with
plaquette variables and make use of
techniques which have become standard practice in the fields of random matrices
and combinatorics. In
\sect{simpleintegral} we derive a generating function which
allows the calculation of integrals of the form
\bea
\int_{{\rm SU}(N)} dU ({\rm Tr} U)^m \overline{({\rm Tr} U)}^n e^{c ({\rm Tr} U + {\rm Tr}U^\dagger)}.
\eea
The work in \sect{morecomplicated} generalises the generating
function of \sect{simpleintegral} allowing the calculation of
more complicated integrals of the form
\bea
\int_{{\rm SU}(N)} dU \left[{\rm Tr}(U^l)\right]^m  e^{c ({\rm Tr} U + {\rm Tr}U^\dagger)}.
\eea

For each integral considered the approach is the same and proceeds as
follows. We
start with a calculation of a  U($N$) integral. For example in
\sect{simpleintegral} we calculate 
\bea
G_{{\rm U}(N)}(c,d) &=& \int_{{\rm U}(N)} dU e^{c {\rm Tr} U + d {\rm Tr} U^\dagger}.
\label{simplegen}
\eea
This is a generalisation of $G_{{\rm U}(N)}(c,c)$, an integral first
calculated by Kogut, Snow and
Stone~\cite{Kogut:1982ez}. We then make use of a result of Brower, Rossi and
Tan~\cite{Brower:1981vt} to extend the U($N$) integral to SU($N$) by
building the restriction, $\det U = 1$ for all $U \in$ SU($N$), into
the integration measure. In this way SU($N$)
generating functions can be obtained  as sums of determinants whose entries
are modified Bessel functions of the first kind.

\subsubsection{A simple integral}
\label{simpleintegral} 
In this section we introduce a useful technique for performing SU($N$)
integrals. We start with the U($N$) integral of \eqn{simplegen} and
calculate it using a technique which has become standard practice in the
study of random matrices and combinatorics.

Since the Haar measure is left and right invariant (see
\eqn{LRinvariance}) we can diagonalise $U$ inside the integral as 
\bea
 U = V \left(\begin{array}{cccc} e^{i\phi_0} & 0 &\cdots & 0\\
                                 0 & e^{i \phi_1}&       & \vdots \\
                                \vdots & & \ddots &   \\
                                0 & \cdots & & e^{i\phi_N}
                                 \end{array}\right) 
V^\dagger.
\eea
In terms of the set of variables $\{\phi_k\}_{k=1}^{N}$ the Haar
measure factors as $dU = d\mu(\phi) d V$~\cite{Kogut:1982ez}. Since
the integrand is independent of $V$, the $V$ integral can be carried
out trivially using the normalisation of the Haar measure given by 
\eqn{normalisation}.

Making use of the Weyl parameterisation for U($N$)~\cite{Weyl:1946},
\bea
d\mu(\phi) &=& \prod_{i=1}^{N} \frac{d\phi_i}{2\pi} |\Delta(\phi)|^2 ,
\label{unparam}
\eea
where $\Delta(\phi)$ is the Vandermonde determinant, with implicit
sums over repeated indices understood,
\bea
\Delta(\phi) &=& \frac{1}{\sqrt{N!}} \varepsilon_{i_1 i_2\cdots i_N} e^{i
\phi_1(N-i_1)}e^{i
\phi_2(N-i_2)} \cdots  e^{i
\phi_N(N-i_N)},
\label{vandermonde}
\eea
 we can express the
U($N$) generating function as follows
\bea
G_{{\rm U}(N)}(c,d) &=& \int_0^{2
\pi}\frac{d\phi_1}{2\pi}\cdots\int_0^{2 \pi}\frac{d\phi_N}{2\pi}
\exp\left[\sum_{i=1}^{N}(c e^{i\phi_i}+d e^{-i\phi_i})\right]|\Delta(\phi)|^2.\label{Weyl}
\eea
In \eqn{vandermonde}, 
$\varepsilon_{i_1\ldots i_n}$ is the totally antisymmetric
Levi-Civita tensor defined to be 1 if $\{i_1,\ldots,i_n\}$ is an even
permutation of $\{1,2,\ldots,n\}$, $-1$ if it is an odd permutation
and 0 otherwise (i.e. if an index is repeated).
Substituting \eqn{vandermonde} in \eqn{Weyl} gives,
\bea
G_{{\rm U}(N)}(c,d) &=& 
\frac{1}{N!} 
\varepsilon_{i_1 i_2\ldots i_N}\varepsilon_{j_1 j_2\ldots j_N}\prod_{k=1}^N \int_0^{2
\pi}\frac{d\phi_k}{2\pi}\exp\left[i(j_k-i_k)\phi_k+ c e^{i \phi_k} +d
e^{-i \phi_k} \right]. \label{U(N)integral}
\eea
To simplify this further we need an expression for the integral,
\bea
g_n(c,d) &=& \int_0^{2\pi} \frac{dx}{2\pi} \exp(i n x + c e^{i x} + d
e^{-i x}),
\eea
which is easily handled by expanding the integrand in Taylor series in
$c$ and $d$,
\bea
g_n(c,d) &=&
\sum_{k=0}^\infty \sum_{l=0}^\infty \frac{c^k d^l}{k! l!}\int
\frac{dx}{2\pi} e^{i x\left(k-l + n \right)} \nn\\
&=& \sum_{k=0}^\infty  \frac{c^k d^{k+n}}{k! (k+n)!} \nn\\
&=& \left(\frac{d}{c}\right)^{n/2} I_n\left(2 \sqrt{c d}\right).
\label{neededint}
\eea
Making use of \eqn{neededint} in \eqn{U(N)integral} gives an
expression for $G_{{\rm U}(N)}(c,d)$ as a Toeplitz
determinant (a determinant of a
matrix whose $(i,j)$-th entry depends only on $j-i$),
\bea
G_{{\rm U}(N)}(c,d) &=& 
\frac{1}{N!} 
\varepsilon_{i_1 i_2\ldots i_N}\varepsilon_{j_1 j_2\ldots j_N}
\prod_{k=1}^N g_{j_k-i_k}(c,d) \nn\\
&=&
\frac{1}{N!} 
\varepsilon_{i_1 i_2\ldots i_N}\varepsilon_{j_1 j_2\ldots j_N}
 \left(\frac{d}{c}\right)^{\sum_{l=0}^N (i_l - j_l)/2}
\prod_{k=1}^N I_{j_k-i_k}\left(2\sqrt{c d}\right) \nn\\
&=& \det\left[ I_{j-i}\left(2\sqrt{c d}\right)\right]_{1\le i,j\le N}.
\eea 
Here the quantities inside the determinant are to be interpreted as
the $(i,j)$-th entry of an $N\times N$ matrix. 
Now to calculate the corresponding 
SU($N$) result the restriction $\det U = 1$,
which is equivalent to $\sum_{k=1}^N \phi_k = 0\,{\rm
mod}\, 2\pi$ in terms of the $\phi_k$ variables, must
be built into the integration measure. 
To do this we follow Brower,
Rossi and Tan~\cite{Brower:1981vt} and incorporate the following delta
function in the integrand of \eqn{Weyl}:
\bea
2\pi \delta\left(\sum_{k=1}^N \phi_k - 0\,{\rm mod}\, 2\pi\right) &=&
\sum_{m=-\infty}^{\infty} 2 \pi \delta\left(\sum_{k=1}^N \phi_k - 2\pi
m\right).
\eea 
This is most conveniently introduced into the integral via its Fourier
transform,
\bea
\sum_{m=-\infty}^{\infty} \exp\left(i m \sum_{k=1}^N \phi_k\right).
\label{fourtran}
\eea
To obtain the SU($N$) integral from the corresponding U($N$) result the
modification is therefore trivial. Including \eqn{fourtran} in the
integrand of \eqn{Weyl} leads to the general SU($N$) result,
\bea
G_{{\rm SU}(N)}(c,d) 
&=& \int_{{\rm SU}(N)} dU e^{c {\rm Tr} U + d {\rm
Tr} U^\dagger}\nn\\
&=& \sum_{m=-\infty}^{\infty} \det
\left[g_{m+j-i}(c,d)\right]_{1\le i,j\le N}.
\label{simplealmost}
\eea
This expression can be manipulated to factor the $d/c$ dependence out of the
determinant as follows,
\bea
\det \left[\left(\frac{d}{c}\right)^{(l+j-i)/2}I_{l+j-i}\left(2\sqrt{c
d}\right)\right]_{1\le i,j\le N} \hspace{-0.5cm}&=&
\frac{1}{N!}\varepsilon_{i_1 i_2\ldots i_N}\varepsilon_{j_1 j_2\ldots
j_N}\left(\frac{d}{c}\right)^{l N/2 +\sum_k(j_k-i_k)/2}\nn\\
&&\times
\prod_{m=1}^{N}I_{l+j_m-i_m}\left(2\sqrt{c d}\right) \nn\\
&=& \left(\frac{d}{c}\right)^{l N/2} 
\det \left[I_{l+j-i}\left(2\sqrt{c d}\right)\right]_{1\le i,j\le N}.
\eea
Making use of this result in \eqn{simplealmost} leads to the SU($N$)
generating function
\bea
G_{{\rm SU}(N)}(c,d) &=&
\sum_{l=-\infty}^{\infty} \left(\frac{d}{c}\right)^{l N/2}
\det \left[ I_{l+j-i}\left(2\sqrt{cd}\right)\right]_{i\le i,j \le N}.
\label{coolsum}
\eea
For the case of SU(2) we can show that this reduces to the standard
result of Arisue given by \eqn{su2gen}. To do this we need the
recurrence relation for modified Bessel functions of the first kind,
\bea
I_{n-1}(x)-I_{n+1}(x) = \frac{2n}{x} I_n(x).
\label{irecurr}
\eea
Recall that for SU(2) the Mandelstam constraint is $\Tr U = \Tr
U^\dagger$, so the case $G_{{\rm SU}(2)}(c,c)$ can be considered without
loss of generality;
\bea
G_{{\rm SU}(2)}(c,c) &=& \sum_{l=-\infty}^{\infty}
\left[I_l(2c)^2-I_{l-1}(2c)I_{l+1}(2c) \right] \nn\\
&=& I_0(4c) - I_2(4c). 
\eea
Here we have used the standard addition
formula~\cite{Gradshteyn:1994} for modified Bessel functions. Employing the recurrence relation of \eqn{irecurr} gives
\bea
G_{{\rm SU}(2)}(c,c) &=& \frac{1}{2 c} I_1(4 c),
\label{su2simple}
\eea
which is the standard result of Arisue given by \eqn{su2gen}.

With an analytic form for SU($N$) in hand we can attempt to find
simpler expressions for $G_{\rm{SU}(N)}(c,d)$ analogous to \eqn{su2simple}.
To our knowledge no general formulas are available for the simplification
of the determinants appearing in \eqn{coolsum}. Without such formulas
we can resort to the crude method of analysing series expansions and
comparing them with known expansions of closed form expressions. Since
the determinants appearing in the generating function are nothing more
than products of modified Bessel functions we expect that if a closed
form expression for the general SU($N$) generating function exists,
it will involve the generalised hypergeometric function. With this
approach we have limited success. The SU(3) result of \eqn{Y} is
recovered numerically but the SU(4) result does simplify analytically. 

When analysing the series expansion of $G_{\rm{SU}(4)}(c,c)$ we notice that
it takes the form of a generalised hypergeometric function. In
particular we find the following result:  
\bea
G_{{\rm SU}(4)}(c,c) &=& {}_2F_3\left[\begin{array}{c} \frac{3}{2}, \frac{5}{2}
\\ 3 , 4 , 5\end{array}; 16 c^2\right].\label{su4hyper}
\eea
Here the generalised hypergeometric function is defined by
\bea
{}_pF_q\left[ \begin{array}{c} 
a_1, a_2,\ldots ,a_p \\
b_1,b_2,\ldots ,b_q
\end{array}; x\right] = \sum_{k=0}^\infty
\frac{(a_1)_k (a_2)_k\ldots
(a_p)_k}{(b_1)_k (b_2)_k\ldots (b_q)_k} \frac{x^k}{k!},
\eea 
where $(x)_k=x(x+1)\ldots(x+k-1)$ is the rising factorial or 
Pochhammer symbol. 
In addition to \eqn{su4hyper} we find the following results for matrix
elements derived from the SU(4) generating function:
\bea
\langle Z_1 \rangle &\!\!\!=\!\!\!& \frac{c\, {}_2F_3\left[\!\begin{array}{c} \frac{5}{2}, \frac{7}{2}
\\ 4 , 5,  6\end{array}; 16 c^2\right]}{ {}_2F_3\left[\!\begin{array}{c} \frac{3}{2}, \frac{5}{2}
\\ 3 , 4 , 5\end{array}; 16 c^2\right]},
\eea
\bea
\hspace{-1cm}\langle Z_1^2 \rangle &\!\!\!=\!\!\!& \frac{\frac{3}{2}c^2 \, {}_2F_3\left[\!\begin{array}{c} \frac{5}{2}, \frac{7}{2}
\\ 5, 6, 7\end{array}; 16 c^2\right] + \frac{2}{3} c^4\, {}_2F_3\left[\!\begin{array}{c} \frac{7}{2}, \frac{9}{2}
\\ 6, 7, 8\end{array}; 16 c^2\right] +\frac{1}{15} c^6\, {}_2F_3\left[\!\begin{array}{c} \frac{9}{2}, \frac{11}{2}
\\ 7, 8, 9\end{array}; 16 c^2\right] }{ {}_2F_3\left[\!\begin{array}{c} \frac{3}{2}, \frac{5}{2}
\\ 3, 4, 5\end{array}; 16 c^2\right]}
\eea
and 
\bea
\hspace{-1cm}\langle Z_1\bar{Z}_1 \rangle &\!\!\!=\!\!\!& \frac{ {}_2F_3\left[\!\begin{array}{c} \frac{3}{2}, \frac{5}{2}
\\ 3, 5, 6\end{array}; 16 c^2\right] + \frac{4}{3} c^2\, {}_2F_3\left[\!\begin{array}{c} \frac{5}{2}, \frac{7}{2}
\\ 4, 6, 7\end{array}; 16 c^2\right] +\frac{4}{9} c^4\, {}_2F_3\left[\!\begin{array}{c} \frac{7}{2}, \frac{9}{2}
\\ 5, 7, 8\end{array}; 16 c^2\right]
}{ {}_2F_3\left[\!\begin{array}{c} \frac{3}{2}, \frac{5}{2}
\\ 3, 4, 5\end{array}; 16 c^2\right]}.
\eea
We stress that these results are nothing more than observations based
on series expansions. Despite some effort analogous expressions for $N>4$ have not been found.

The generating functions, $G_{{\rm SU}(N)}(c,d)$  and $G_{{\rm
U}(N)}(c,d)$, are not only of interest in Hamiltonian LGT.  
By differentiating \eqn{coolsum} appropriately with respect to $c$
and $d$ and afterwards setting $c$ and $d$ to zero, we obtain the
number of singlets in a given product representation of SU($N$). This
was discussed for the special case of SU(3) in \sect{su3integrals}.  
We now consider the general case in the calculation of
$T_k(n)$; the number of singlets in the SU($k$) product representation,
\bea
\underbrace{({\bf k} \otimes {\bf\bar{k}})\otimes \cdots\otimes
({\bf k} \otimes {\bf \bar{k}})}_{n}.
\eea 
As a group integral $T_k(n)$ is given by
\bea
T_k(n) = \int_{{\rm SU}(k)} d U(|\Tr U |^2  )^n.
\eea
Integrals of this kind are studied in combinatorics, in particular the study of increasing
subsequences of permutations. An increasing subsequence is a sequence 
$i_1<i_2 <\cdots < i_m $ such that $\pi(i_1) < \pi(i_2)< \cdots
<\pi(i_m)$, where $\pi$ is a permutation of $\{1,2,\ldots,k\}$. It has been shown that the number of permutations $\pi$ of
$\{1,2,\ldots,k\}$ such that $\pi$ has no increasing subsequence of
length greater than $n$ is $T_k(n)$ \cite{Rains:1998}. 
In addition it is possible to prove that $T_k(n)$ is the
number of pairs of Young tableaux of size $k$ and maximum height
$n$ via the Schensted correspondence~\cite{vanMoerbeke:2001,Rains:1998}. 

Making use of \eqn{coolsum} we see that only the $l=0$ term
contributes to $T_k(n)$. Letting $x=cd$ we have:
\bea
T_k(n) &=& (\partial_c\partial_d)^n \det \left[I_{j-i}(2\sqrt{c
d})\right]_{1\le i,j \le k}\Bigg|_{c=d=0} \nn\\
&=& n! \frac{d^n}{d x^n} \det
\left[I_{j-i}(2\sqrt{x})\right]_{1\le i,j \le k} \Bigg|_{x=0}.
\eea
Hence the generating function for $T_k(n)$ is given by
\bea
\sum_{n=0}^\infty \frac{T_k(n) x^n}{n!^2} &=& \det \left[I_{j-i}(2\sqrt{x})
\right]_{1\le i,j \le k}, 
\eea 
a result first deduced by Gessel~\cite{Gessel:1990}.
The first few $T_k(n)$ sequences are available as A072131, A072132,
A072133 and A072167 in Sloane's on-line encyclopedia of integer
sequences~\cite{Sloane:OE}.

\subsubsection{A more complicated integral}
\label{morecomplicated}

We now move on to the more complicated integral
\bea
H_m(c,d) &=& \int_{{\rm SU}(N)} dU \exp\left[c ({\rm Tr} U + {\rm Tr}
U^\dagger)+ d{\rm Tr}(U^m)\right]\quad \forall m\in \mathbb{Z}^+ . 
\eea
This integral is of interest as a generating function for the
calculation of integrals such as
\bea
\int_{{\rm SU}(N)} dU {\rm Tr}(U^m) e^{c ({\rm Tr} U + {\rm Tr}U^\dagger)}.
\label{egint}
\eea
For the simple case of SU(3) we can use the Mandelstam constraint, ${\rm Tr}(U^2) =  ({\rm Tr}U)^2-2 {\rm Tr} U^\dagger$, to reduce such
integrals to those obtainable from $G_{{\rm SU}(N)}(c,d)$ . However for higher dimensional gauge
groups not all trace variables can be written in
terms of ${\rm Tr} U$ and  ${\rm Tr} U^\dagger$. For these gauge
groups one must introduce the generating function, $H_m(c,d)$, to
calculate integrals similar to  \eqn{egint}.  

To calculate $H_m(c,d)$ we start with the corresponding U($N$)
generating function $h_m(c,d)$ and follow the procedure of \sect{simpleintegral}
to obtain
\bea
h_m(c,d) &=& \int_{{\rm U}(N)} dU \exp\left[c ({\rm Tr} U + {\rm Tr}
U^\dagger)+ d{\rm Tr}(U^m)\right]\nn\\
&&\hspace{-1.5cm}= \frac{1}{N!} 
\varepsilon_{i_1\ldots i_N}\varepsilon_{j_1\ldots j_N}
\prod_{k=1}^N \int_0^{2
\pi}\frac{d\phi_k}{2\pi}\exp\left[i(j_k-i_k)\phi_k+ 2 c 
\cos\phi_k+ d e^{m i \phi_k}\right].
\label{ungen}
\eea
To proceed we need the following integral,
\bea
\int_{0}^{2\pi} \frac{dx}{2\pi} \exp(i n x + a \cos x + b e^{i m x})
&=&  \sum_{k=0}^{\infty} \frac{b^k}{k!} \int_{0}^{2\pi}
\frac{dx}{2\pi} e^{i (n+m k) x + a \cos x}\nn\\
&=&  \sum_{k=0}^{\infty} \frac{b^k}{k!} I_{n+m k}(a). 
\label{needthis}
\eea
Making use of \eqn{needthis} in \eqn{ungen} leads to the following
expression for the U($N$) generating function, 
\bea
h_m(c,d) &=& \det \left[ \lambda_{m;j-i}(c,d)\right]_{1\le i,j\le N},
\eea
with 
\bea
\lambda_{m;n}(c,d) = \sum_{k=0}^{\infty} \frac{d^k}{k!} I_{n+m k}(2 c).
\label{lam}
\eea
Extending to SU($N$) following the prescription of 
\sect{simpleintegral}, 
we arrive at the corresponding SU($N$) generating function,
\bea
H_m(c,d) &=& \sum_{l=-\infty}^{\infty}\det
\left[\lambda_{m;l+j-i}(c,d) \right]_{1\le i,j\le N} .
\label{coolersum}
\eea
An example of an SU($N$) integral derived from this
generating function is the following:
\bea
\int_{{\rm SU}(N)} dU {\rm Tr}(U^m) e^{c ({\rm Tr} U + {\rm Tr}
U^\dagger)} &=& \frac{\partial H_m(c,d)}{\partial d}\Bigg|_{d=0} \nn\\
&=& \frac{\partial}{\partial
d}\sum_{l=-\infty}^{\infty}\det\left[I_{l+j-i}(2c)+d I_{l+j-i+m}(2c)\right]\Bigg|_{d=0}.
\label{coolintegral}
\eea
Only two terms need to be kept in the $k$-sum of \eqn{lam} here
because higher order powers of $d$ vanish when the derivative with
respect to $d$ is taken and $d$ set to zero.

%

\section{Analytic SU($N$) Calculations}    
\label{analyticSU(N)calcs}
\subsection{Preliminaries}
\label{prelims}
In this section we make use of the analytic results of
\sect{suNintegrals} in variational calculations of glueball
masses. The steps we take are as follows. We first fix the variational
ground state by minimising the vacuum energy density. Having fixed the
ground state we are then free to investigate the 
excited states.

Before fixing the variational ground state we introduce the following 
convenient notation.
We define the general order $a^2$ improved lattice Hamiltonian 
for pure SU($N$) gauge theory with coupling $g^2$ on a lattice with
spacing $a$ by 
\bea
\tilde{\LH}(\kappa,u_0) &=& \frac{g^2}{a}\sum_{\boldx,i}
\Tr\left[(1-\kappa)\LE_i(\boldx)^2 + \frac{\kappa}{u_0^2}
\LE_i(\boldx) U_i(\boldx) \LE_i(\boldx+a \boldsymbol{i})
U^\dagger_i(\boldx)\right]\nn\\
&& + \frac{2N}{a g^2} \sum_{\boldx, i<j}\left\{(1+4\kappa) P_{ij}(\boldx) -  
\frac{\kappa}{2} \left[R_{ij}(\boldx)+R_{ji}(\boldx)\right]\right\},
\label{genham}
\eea
where the plaquette and rectangle operators are given by
\be
\begin{array}{c}\includegraphics{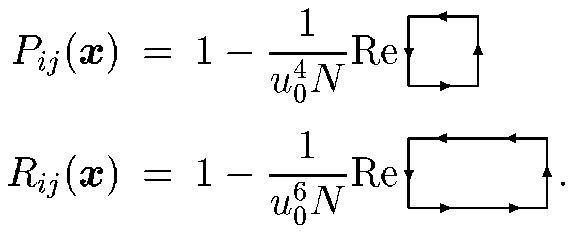}\end{array}
\label{tadplaqrect}
\ee
The simplest lattice Hamiltonians derived in \rcite{Carlsson:2001wp}
 can be expressed 
in terms of $\tilde{\LH}$ as
follows. The Kogut-Susskind~\cite{Kogut:1975ag} 
and ${\cal O}(a^2)$ classically 
improved Hamiltonians~\cite{Carlsson:2001wp} are given by $\tilde{\LH}(0,1)$ 
and $\tilde{\LH}(1/6,1)$ respectively.   
The tadpole improved Hamiltonian~\cite{Carlsson:2001wp} 
is given by $\tilde{\LH}(1/6,u_0)$,
where the mean link $u_0$, which can be expressed in terms of the mean
plaquette, 
\be
\begin{array}{c}\includegraphics{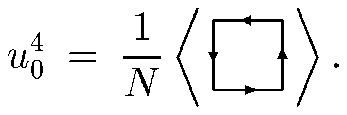}\end{array}
\label{meanlink}
\ee
is defined self-consistently as a function of $\beta = N/g^2$ as
described in \sect{fixingtheintroduction}. 

With this notation the vacuum energy density is given by
\be
\begin{array}{c}\includegraphics{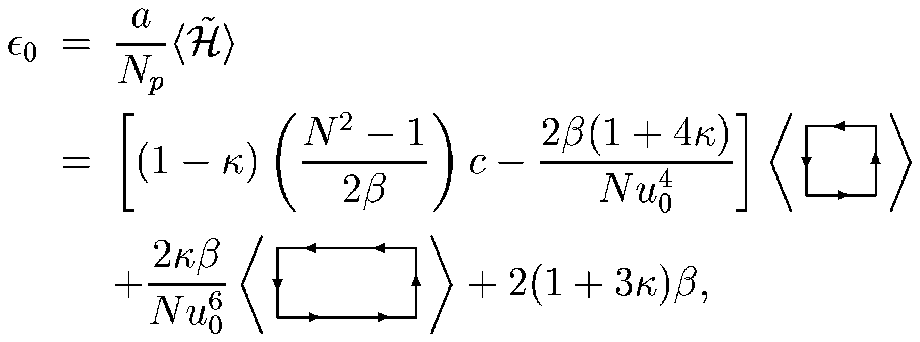}\end{array}
\label{epsilon}
\ee
where $N_p$ is the number of
plaquettes on the lattice and the expectation values, as usual, are
taken with respect to the one plaquette
trial state defined in \eqn{oneplaquette}. The variational parameter,
$c$, is fixed as a function of
$\beta$ by minimising the vacuum energy density. For the calculation
of the expectation values we use the generating functions of
\sect{suNintegrals} with all infinite sums truncated. 
Once the variational parameter is fixed the trial state is completely
defined as a function of $\beta$.

\section{Fixing The Variational Trial State}
\label{fixingthevariational}
\subsection{Introduction}
\label{fixingtheintroduction}
In this section we fix the SU($N$) trial state for $2\le N\le 5$, 
making use of the generating functions derived in
\sect{suNintegrals}. 
These generating functions allow the
analytic calculation of the plaquette and rectangle expectation values
appearing \eqn{epsilon}. The approach
we take is as follows. For the Kogut-Susskind and
classically improved cases, we simply minimise $\epsilon_0$ for a given
value of $\beta$. The value of $c$ at which $\epsilon_0$ takes its minimum
defines $c$ as a function of $\beta$. The tadpole improved case is
more complicated because the mean plaquette depends
on the variational state, which is determined by minimising the energy
density. The energy density however, depends on the mean plaquette. 
Such interdependence suggests the use of an iterative procedure for
the calculation of the tadpole improved 
energy density. The approach we adopt is as
follows. For a given $\beta$
and starting value of $u_0$ we minimise the energy density of
\eqn{epsilon} to fix the
variational state $|\phi_0\rangle$. We then calculate a new mean
plaquette value using this trial state and substitute in
\eqn{epsilon} to obtain a new 
expression for the energy density. This process is iterated until 
convergence is achieved, typically between five and ten iterations.

\subsection{Results}
\begin{figure*}
\centering
\subfigure[SU(2)] 
                     {
                         \label{esu2}
                         \includegraphics[width=7cm]{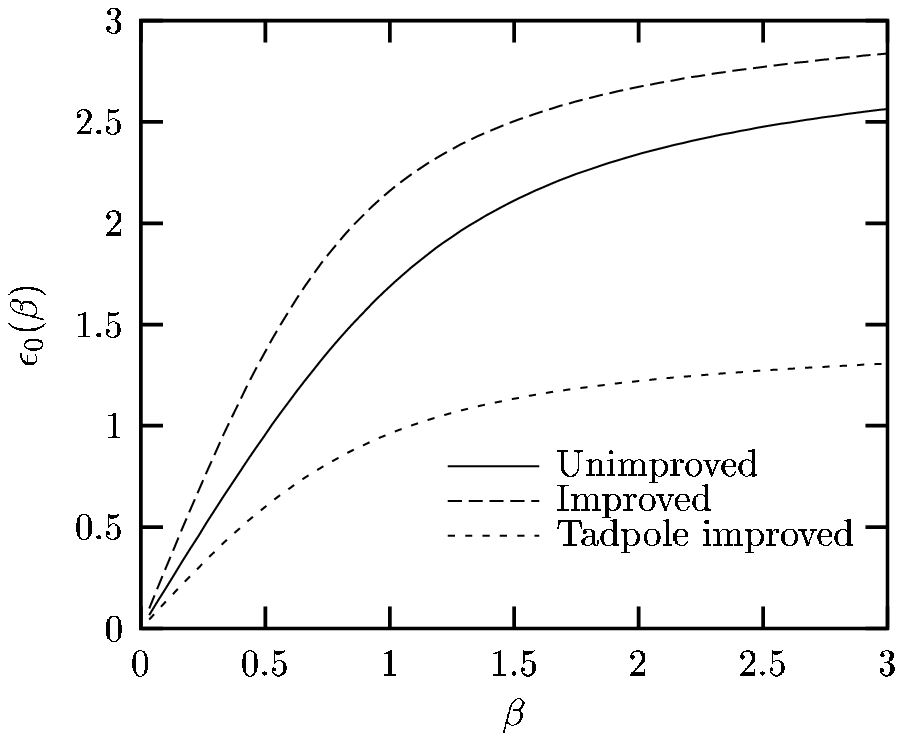}
                     }                   
 \subfigure[SU(3)] 
                     {
                         \label{esu3}
                         \includegraphics[width=7cm]{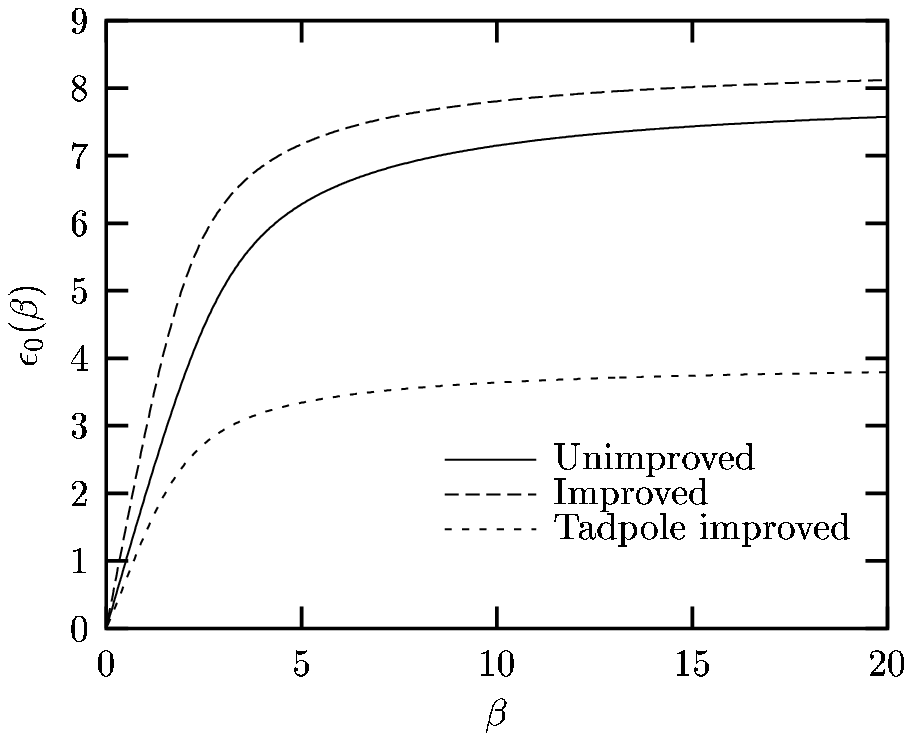}
                     }\\
        \subfigure[SU(4)] 
                     {
                         \label{esu4}
                         \includegraphics[width=7cm]{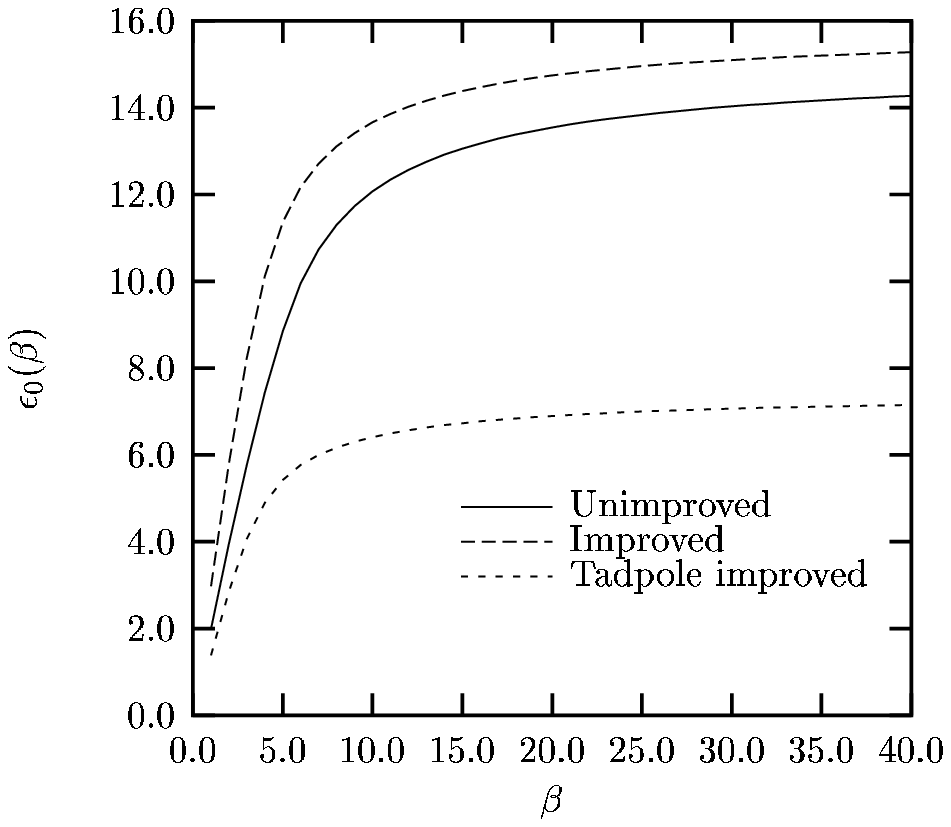}
                     }
\subfigure[SU(5)] 
                     {
                         \label{esu5}
                         \includegraphics[width=7cm]{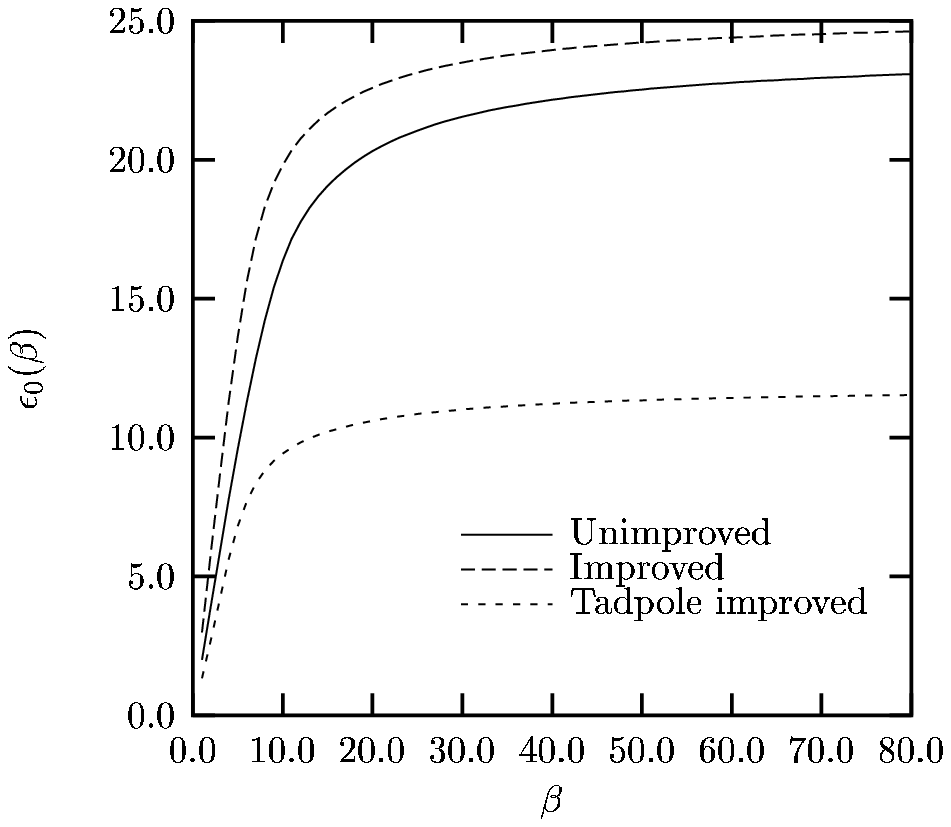}
                     }
\caption{Analytic calculation of the 2+1 dimensional 
unimproved, improved and tadpole
                         improved vacuum energy density in units of
                         $1/(a N_p)$ for SU(2), SU(3), SU(4) and SU(5).}
\label{edens}
\end{figure*}

\begin{figure*}
\centering
\subfigure[SU(2)] 
                     {
                         \label{csu2}
                       \includegraphics[width=7cm]{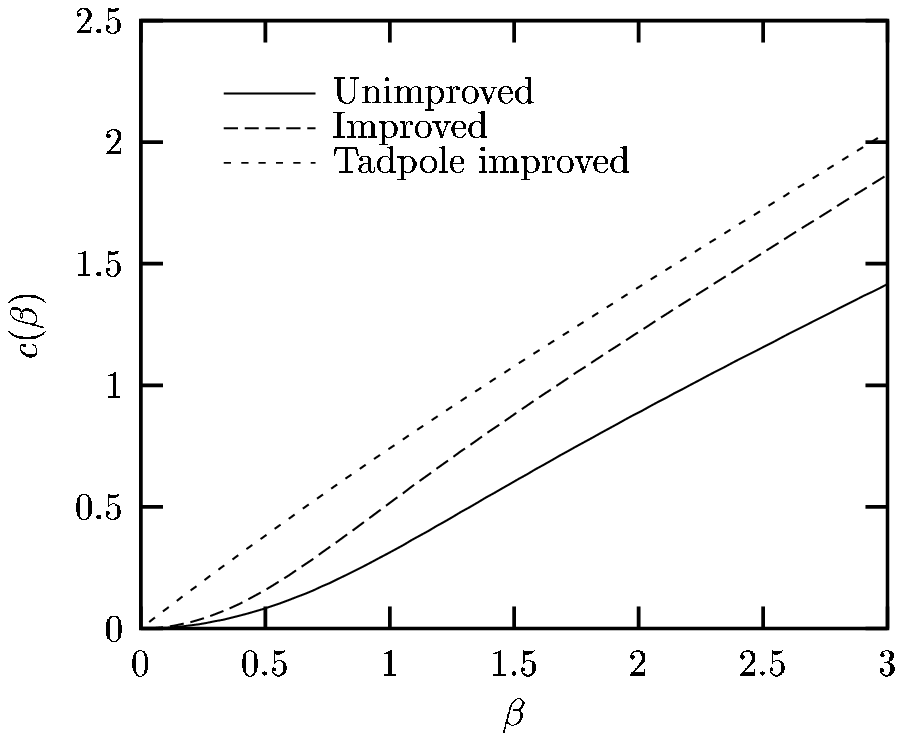}
                     } \hspace{0.25cm}                   
 \subfigure[SU(3)] 
                     {
                         \label{csu3}
                         \includegraphics[width=7cm]{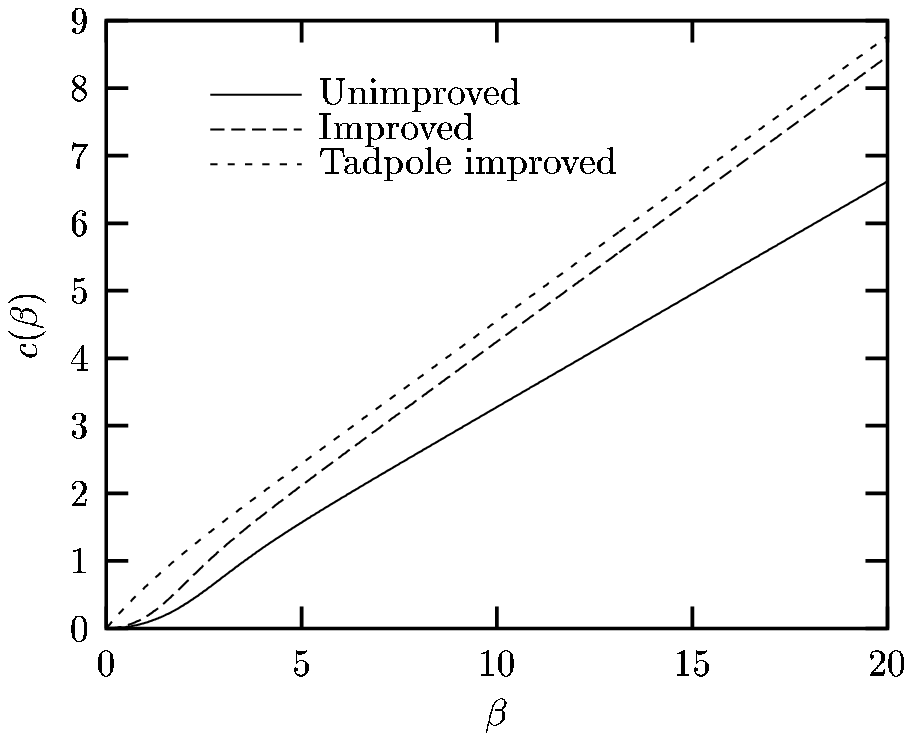}
                     }
\subfigure[SU(4)] 
                     {
                         \label{csu4}
                       \includegraphics[width=7cm]{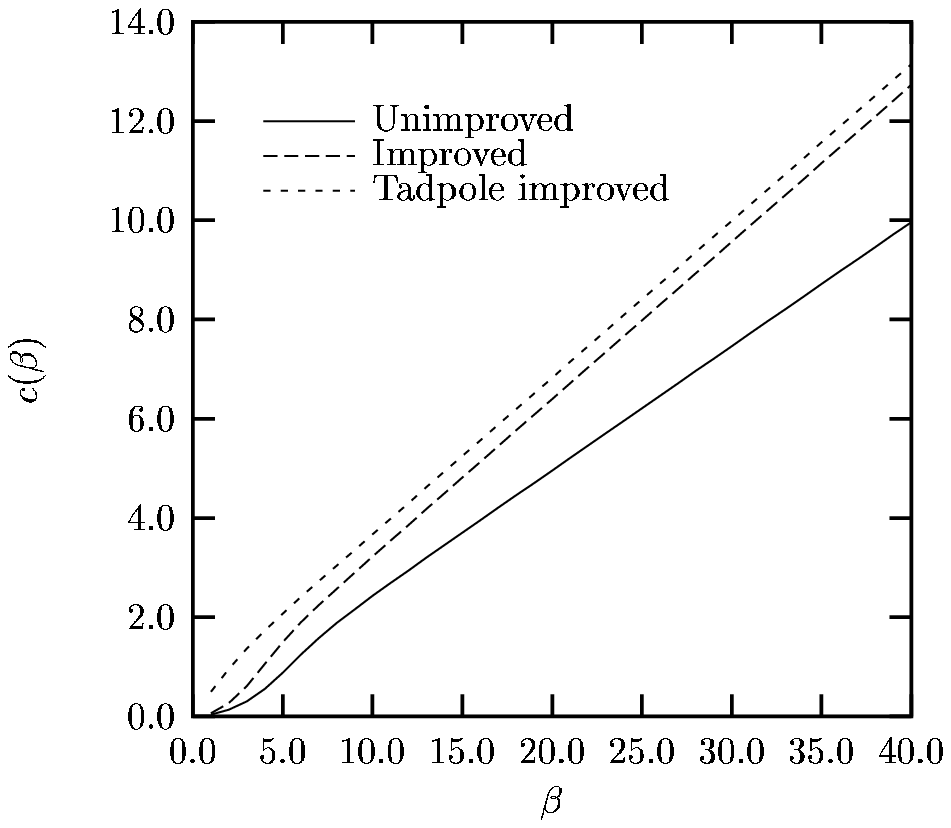}
                     } \hspace{0.25cm}                   
 \subfigure[SU(5)] 
                     {
                         \label{csu5}
                         \includegraphics[width=7cm]{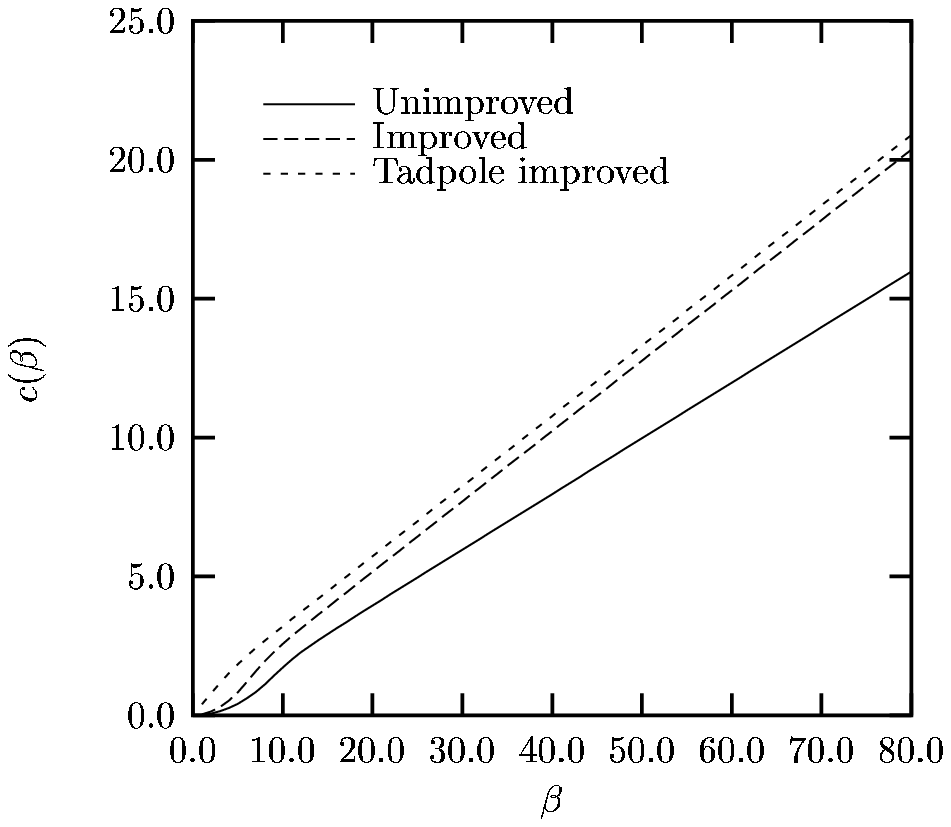}
                     }
\caption{Analytic calculation of the unimproved, improved and tadpole
                         improved variational parameter in 2+1
                         dimensions for SU(2), SU(3), SU(4) and SU(5).}
\label{cofbeta}
\end{figure*}

The
results of the Kogut-Susskind, classically improved and tadpole
improved  SU(2), SU(3), SU(4) and SU(5) vacuum energy density calculations are shown
in \fig{edens}. The corresponding variational parameters $c(\beta)$
 are shown in \fig{cofbeta}. For SU(3) the generating function of
\eqn{Y} is used to calculate the required plaquette and rectangle
expectation values.  The generating function
of \eqn{coolsum} is used for SU(4) and SU(5).

The familiar strong and weak coupling behavior from variational
calculations is observed in each case.
The differing gradients for the improved and Kogut-Susskind $c(\beta)$ for a given $N$ in the weak coupling limit highlight the fact that when using an improved Hamiltonian one is using a different renormalisation scheme to the unimproved case.


\subsection{Dependence on truncation}
\label{dependenceontruncation}
\begin{figure}
\centering

\includegraphics[width=7cm]{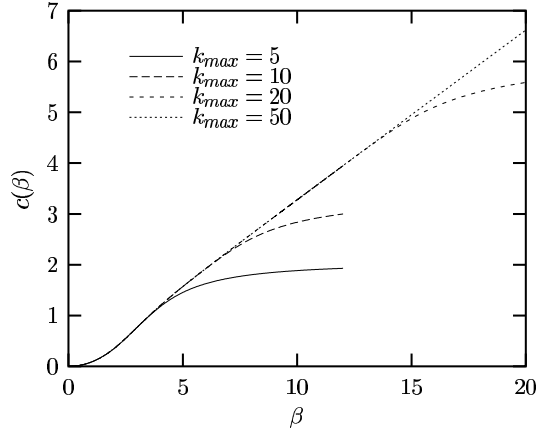}

\caption{Analytic calculation of the unimproved SU(3) variational parameter in 2+1 dimensions, truncating the $k$-sum of ${\cal Y}(c,d)$ at $k=k_{max}$.} 
\label{cconvergence}  
\end{figure}

In practice the $k$-sum appearing in the SU(3) generating function is
truncated at a maximum value $k_{max}$.  The dependence of the variational parameter on various truncations of the $k$-sum 
is shown in \fig{cconvergence}. We see that convergence is achieved up
to $\beta \approx 13$ when keeping 20 terms. Further calculations show
that when keeping 50 terms convergence up to $\beta \approx 30$ is
achieved.

\begin{figure*}
\centering

\subfigure[SU(3)] 
                     {
                         \label{detmaxsu3}
                       \includegraphics[width=7cm]{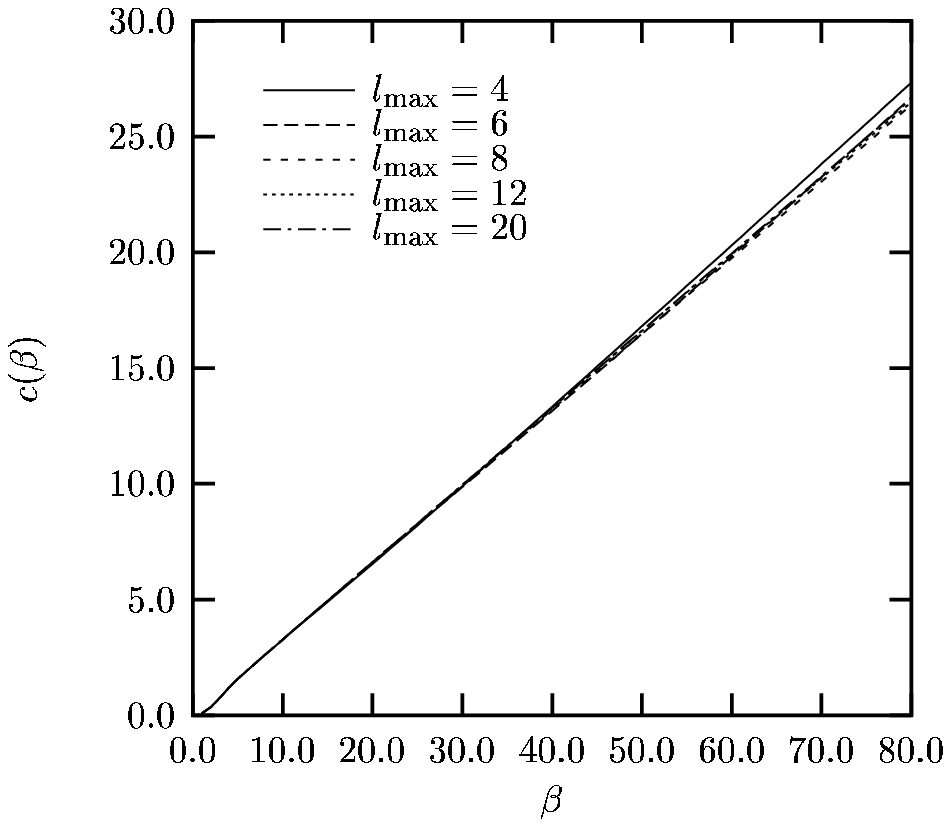}
                     } \hspace{0.25cm}                   
\subfigure[SU(4)] 
                     {
                         \label{detmaxsu4}
                         \includegraphics[width=7cm]{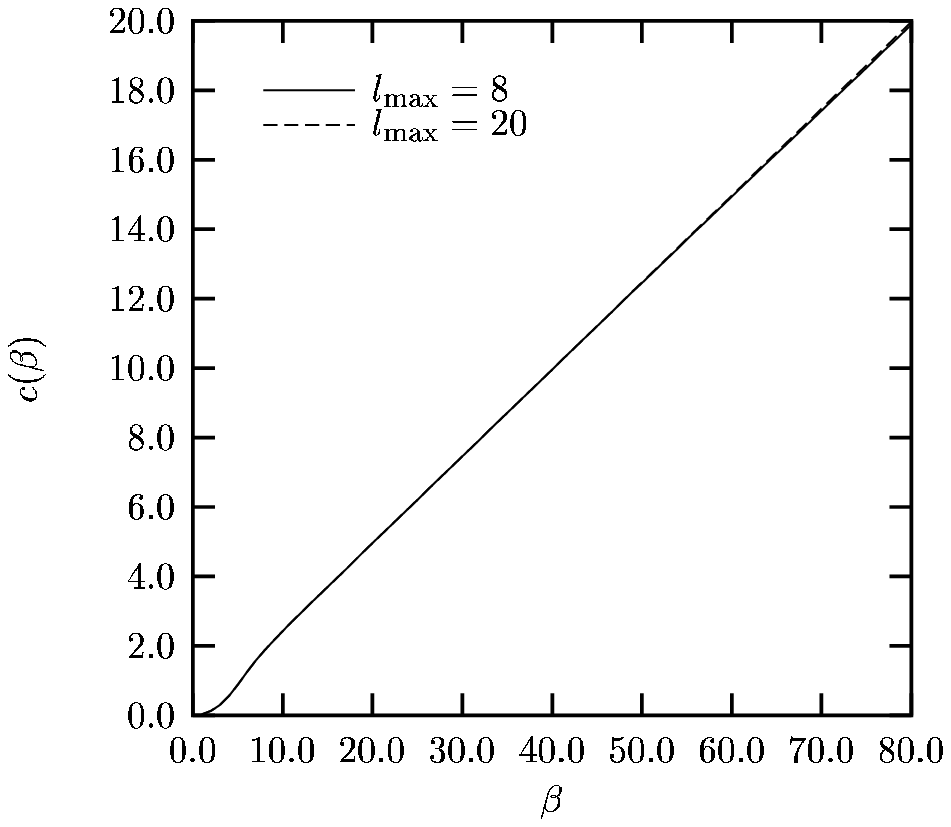}
                     }

\caption{Analytic calculation of the unimproved SU(3) and SU(4)
                         variational parameters in 2+1 dimensions,
                         truncating the $l$-sum of \eqn{coolsum} at
                         $l=\pm l_{max}$.} 
\label{detmaxconvergence}  
\end{figure*}

The $l$-sum appearing in the general SU($N$) generating function of
\eqn{coolsum} is also truncated in practice. We replace the infinite
sum over $l$ by a sum from $-l_{\rm max}$ to $l_{\rm max}$.  The dependence of
the SU(3) and SU(4) variational parameters on $l_{\rm max}$ is shown
in \fig{detmaxconvergence}. From the graphs we see that convergence is achieved
quickly as $l_{\rm max}$ increases for both SU(3) and SU(4). The
results for $l_{\rm max}\ge 8$ are barely distinguishable up to $\beta =80$ with the scale used in the plots.  

\section{Lattice Specific Heat}
\label{specheats-1}
\begin{figure*}
\centering
\subfigure[SU(2)] 
                     {
                         \label{cvsu2}
                         \includegraphics[width=7cm]{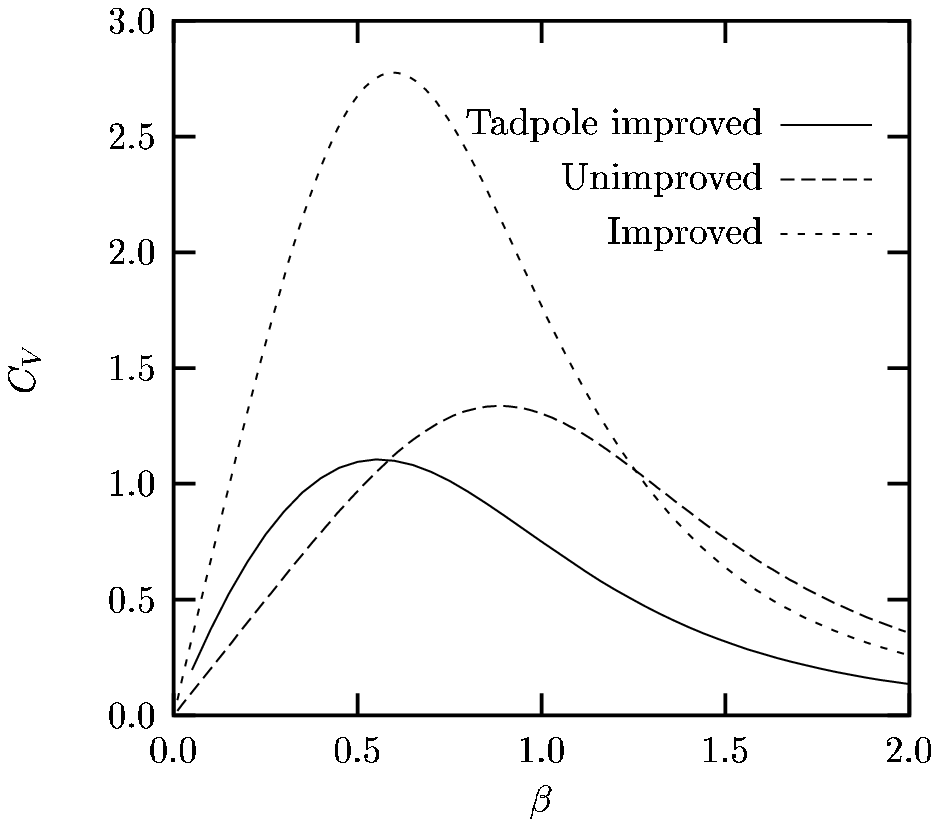}
                     }                   
 \subfigure[SU(3)] 
                     {
                         \label{cvsu3}
                         \includegraphics[width=7cm]{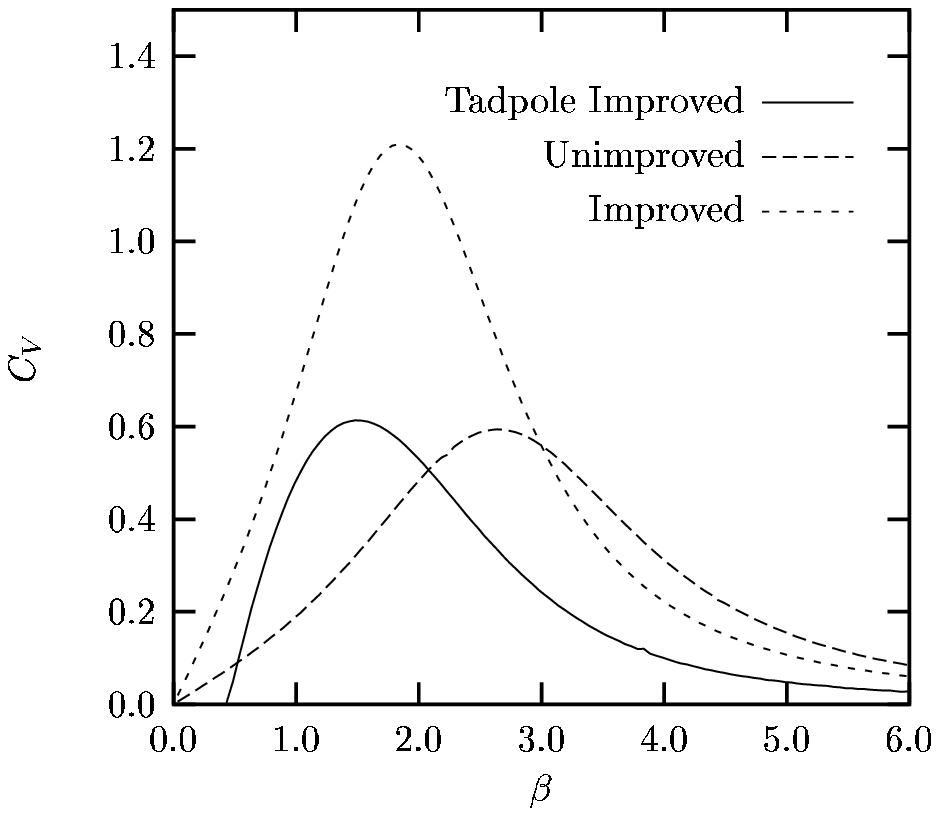}
                     } \\
\subfigure[SU(4)] 
                     {
                         \label{cvsu4}
                         \includegraphics[width=7cm]{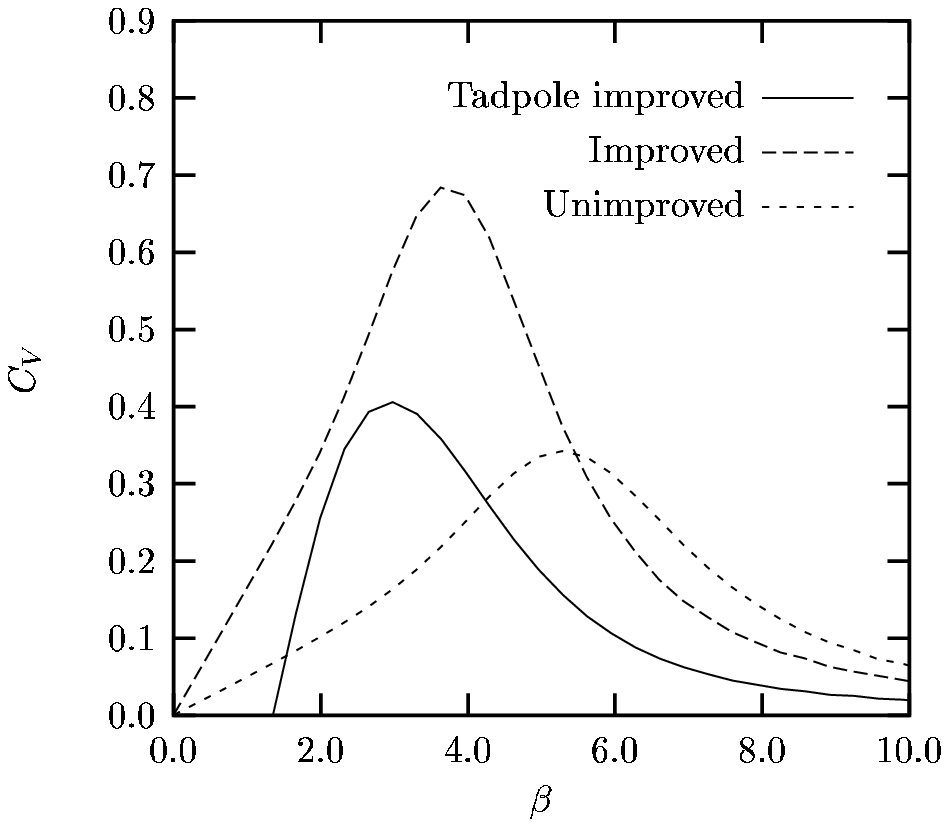}
                     }                   
 \subfigure[SU(5)] 
                     {
                         \label{cvsu5}
                         \includegraphics[width=7cm]{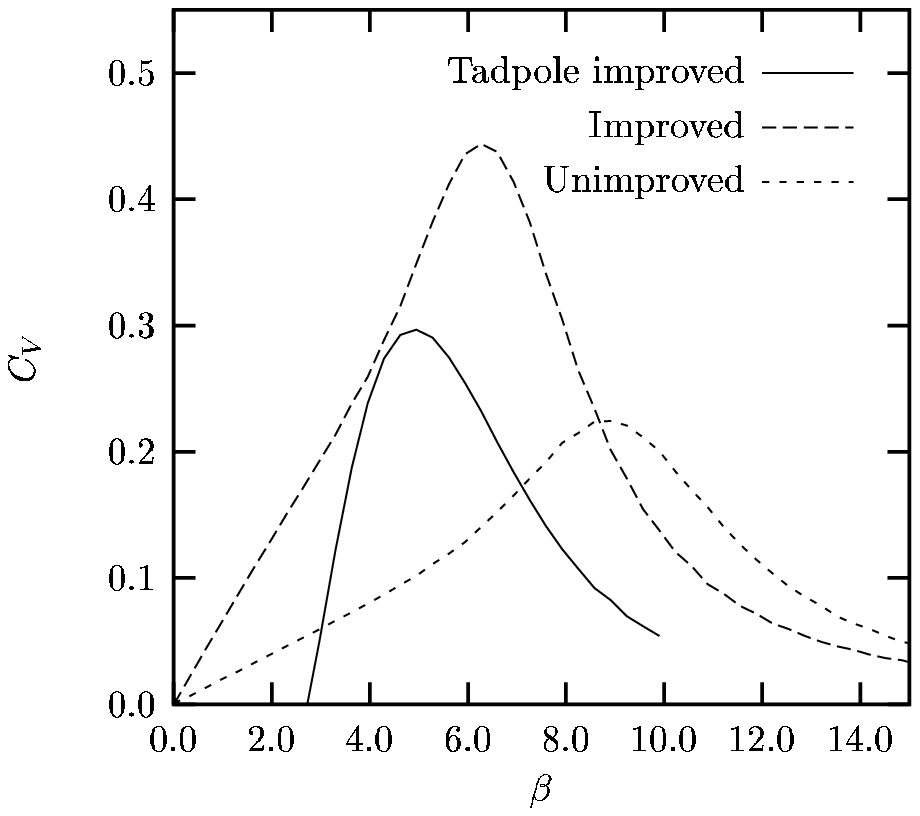}
                     }
\caption{The unimproved, improved and tadpole improved specific heat
                         in 2+1 dimensions for SU(2), SU(3), SU(4) and
                         SU(5).}
\label{cv}
\end{figure*}

In addition to the vacuum energy density we can also calculate the 
lattice specific heat
\bea
C_V = -\frac{\partial^2 \epsilon_0}{\partial \beta ^2}.
\label{latticespecificheat}
\eea
The results for SU(2), SU(3), SU(4) and SU(5) are shown in
\fig{cv}. The SU(2) and SU(3) results are calculated with the aid of
\eqns{su2gen}{Y} with the $k$-sum of \eqn{Y} truncated at $k_{max}=50$.  The SU(4) and SU(5) results are obtained using
\eqn{coolsum} to calculate the required matrix elements. For these
cases the infinite
$l$-sum is truncated at $l_{\rm max}=4$, for which the generating function
has converged on the range of couplings used. We recall that the
location of the peak
indicates the region of transition from strong to weak coupling~\cite{Horn:1985ax}. For an improved
calculation one would expect the peak to be located at a smaller
$\beta$ (corresponding to a larger coupling) than for the equivalent
unimproved calculation. We see that this is indeed the case for each
example, with the tadpole improved Hamiltonian demonstrating the largest degree of improvement.    

\section{Mass Gaps}
\label{massgaps}

\subsection{Introduction}
Having fixed the one-plaquette vacuum wave function, in this section we turn to investigating excited states.
Our aim is to calculate the lowest lying energy eigenstates
of the Hamiltonians described by \eqn{genham} for SU($N$) with $2\le N \le 5$.

We follow Arisue~\cite{Arisue:1990wv} and expand the excited state
$|\phi_1\rangle$ in the basis consisting of all rectangular Wilson
loops 
$\{|n,m\rangle\}_{n,m=1}^{L_{max}}= \{|l\rangle\}_{l=1}^{L_{max}^2}$ that fit in a given square whose side length $L_{max}$
defines the order of the calculation. Enumerating the possible
overlaps between rectangular loops is relatively simple and so a basis
consisting of rectangular loops is an ideal starting point. However,
for an accurate picture of the glueball spectrum we will need to
extend the rectangular basis to include additional smaller area
loops. Without such small area nonrectangular loops, it is possible
that some of the lowest mass states will not appear in the variational
calculation described here.  In order to ensure the orthogonality of $|\phi_0\rangle$ and $|\phi_1\rangle$ we parameterise the excited state as follows
\bea
|\phi_1\rangle &=& \sum_{n,m=1}^{L_{max}} s_{l}|n,m\rangle= \sum_{l=1}^{L_{max}^2} s_{l}|l\rangle, 
\eea
with,
\bea
 |l\rangle &=&  \sum_{\boldx} \left[W_l(\boldx) - \langle W_l(\boldx) 
\rangle\right]|\phi_0\rangle.
\eea
Here $\langle W_l(\boldx) \rangle$ is the expectation value of
$W_l(\boldx)$ with respect to the ground state $|\phi_0\rangle$ and
the convenient label $l=(n-1)L_{max}+m$ has been defined to label the
$n\times m$ rectangular state, $|n,m\rangle$.  We define the particular form of
$W_l(\boldx)$ to reflect the symmetry of the sector we wish to
consider.  For SU($N$) we take $W_l(\boldx)= \Tr[w_l(\boldx)\pm
w^\dagger_l(\boldx)]$ for the symmetric ($0^{++}$) and antisymmetric
($0^{--}$) sectors.  To avoid over-decorated equations, the particular 
$W_{l}(\boldx)$ in use is to be deduced from the context.
Here $w_{l}(\boldx)$ is the rectangular Wilson loop joining
the lattice sites $\boldx$, $\boldx+na\boldsymbol{i}$,
$\boldx+na\boldsymbol{i}+ma\boldsymbol{j}$ and
$\boldx+na\boldsymbol{j}$, with
\bea
n = \left[ \frac{l-1}{L_{max}}\right]+1 \quad{\rm and}\quad m = l- L_{max}\left[ \frac{l-1}{L_{max}}\right] .
\eea
Here $[k]$ denotes the integer part of $k$.
In order to calculate excited state energies we minimise the mass gap (the difference between the excited state and ground state energies) over the basis defined by a particular order $L_{max}$. To do this we again follow Arisue~\cite{Arisue:1990wv} and define the matrices
\bea
N_{l l'} &=& \frac{1}{N_p}\langle l|\tilde{H} - E_0|l'\rangle ,
\label{Nl'l}
\eea
where $E_0$ is the ground state energy, and
\bea
D_{l l'} &=& \frac{1}{N_p}\langle l|l'\rangle 
= \sum_{\boldx}\left[\langle W^\dagger_l(\boldx) W_{l'}(\boldsymbol{0})
\rangle  - \langle W_l(\boldx) \rangle^\ast 
\langle W_{l'}(\boldsymbol{0}) \rangle\right].     
\label{d}
\eea 
Extending the calculation to the general improved Hamiltonian
$\tilde{H}(\kappa,u_0)$ following \rcite{Carlsson:2002ss} gives
\bea
N_{l l'} &=& -\frac{g^2}{2a}\sum_{i,\boldx}\sum_{\boldx'}\Bigg\{
(1-\kappa) \left\langle \left[E^\alpha_i(\boldx),W^\dagger_{l}(\boldx') \right]\left[E^\alpha_i(\boldx),W_{l'}(\boldsymbol{0})\right]\right\rangle \nn\\
&& \hspace{2cm}+ \frac{\kappa}{u_0^2}\left\langle \left[E^\alpha_i(\boldx),W^\dagger_{l}(\boldx')\right]\left[\tilde{E}^\alpha_i(\boldx+a\boldsymbol{i}),W_{l'}(\boldsymbol{0})\right]\right\rangle\Bigg\}.
\label{N}
\eea 
To minimise the mass gap over a basis of a given order we make use of following
 diagonalisation technique~\cite{Tonkin:1987nh}. We first diagonalise
$D$, with 
\bea
S^\dagger D S = V^{2}, \label{diag}
\eea
where $V$ is diagonal.
The $n$-th lowest eigenvalue of the modified Hamiltonian
\bea
H' = V S^\dagger N S V , 
\eea  
then gives an estimate for the mass gap corresponding to the 
$n$-th lowest eigenvalue of the Hamiltonian, $\Delta m_n$.

\subsection{Classification of states}
\label{classificationofstates}

States constructed from only gluon degrees of freedom can be
classified in the continuum by their $J^{PC}$ quantum numbers. 
The assignment of $P$ and $C$ quantum  numbers is straight forward~\cite{Teper:1998te}. Care must be taken, however, when 
building states with particular continuum spins on the
lattice. Difficulties arise when the continuous rotation group of the
continuum is broken down to the group of lattice rotations. The most
serious difficulty to arise is an ambiguity in the assignment of
continuum spins to states built from lattice operators~\cite{Teper:1998te}.
To give a specific example,
suppose we construct a wave function, $|\phi \rangle$, on the lattice  
 with lattice spin $J=0$; a
state built from Wilson loops which are unchanged by rotations of 
$n \pi/2$ for all integers $n$. As explained in \rcite{Teper:1998te}, 
this state  is not
a pure $J=0$ state; it also contains $J=4,8,\ldots $ states. Using a
variational approach we can obtain estimates of the lowest energy
 eigenvalues of states with lattice spin $J=0$. When the continuum
limit is taken we obtain estimates of the lowest continuum energy
eigenvalues for the states with spin $0,4,8,\ldots$. 

In the Lagrangian approach it is possible to suppress the unwanted
spin $J\pm 4,J\pm 8,\ldots $ states in a given spin $J$ calculation. By
``smearing'' links, one can confidently construct lattice states which
do not couple with the unwanted higher spin continuum states, at least for the
lowest energy eigenvalues~\cite{Teper:1998te}. The technique of
``smearing'' links has not to our knowledge been applied in the
Hamiltonian approach. 

 Another way to clarify ambiguities in spin
assignment is to attempt to construct a $J=4$ state on the lattice, devoid
of $J=0$ contributions, and similarly a $J=0$ state devoid of unwanted
$J=4$ contributions. The construction of an exact $J=4$ state is
impossible on a square lattice due to the unavailability of $\pi/4$
rotations. One can however attempt to construct states that are
approximately symmetric under
rotations by $\pi/4$.  In the Lagrangian approach in 2+1 dimensions, it has been demonstrated, for the simple case of SU(2),
 that such states can be chosen on a square
lattice and that the approximate $\pi/4$ symmetry becomes
exact in the continuum limit~\cite{Johnson:1998ev}. 
This technique is readily applicable in
the Hamiltonian approach but has not yet been attempted. 

Thus in our Hamiltonian calculation, which incorporates only
rectangular states,  a lattice spin $J$ state will
correspond to a continuum state with spin $J, J\pm 4, J\pm 8,\ldots $. Using a variational approach we can obtain estimates of the lowest
mass states in the continuum with these spin values. To improve the
spin identification in the continuum beyond modulo $4$ requires more
work.
It will prove interesting to
compare the masses calculated here to that of
Teper~\cite{Teper:1998te} 
who has been careful to identify continuum spins correctly, at least
for the lowest mass excitations.

\subsection{Calculating matrix elements}

Having described the minimisation process
we now detail the calculation of
 the matrix elements $N_{l l'}$ and $D_{l l'}$. Our aim is to reduce $N_{ll'}$ and $D_{ll'}$ to polynomials of one plaquette matrix elements.
This, again has been done for the case of SU(2) by
Arisue~\cite{Arisue:1990wv}. Here we retrace his calculations for the
general case of SU($N$) and extend them to incorporate improved Hamiltonians. We start with $D_{l l'}$.

Taking elementary plaquettes as our independent variables, it is easy to show that the only non-zero contributions to $D_{l l'}$ occur when the rectangles $l$ and $l'$ overlap. As an example of a contribution to $D_{l l'}$, consider $\Delta D_{l l'}$; the case where $N_{l\cap l'}$ plaquettes are contained by both rectangles (these are the overlap plaquettes) and $N_{l}$ plaquettes are contained by the rectangle $l$.
In order to calculate such matrix elements we rely heavily on the orthogonality properties of SU$(N)$ characters. We are interested in calculating SU$(N)$ 
integrals of the form
\bea
\int dU_p e^{S(U_p)}\chi_r(U_p V), 
\label{interest}
\eea 
where $U_p$ is a SU$(N)$ plaquette variable and $V$ is a product of
any number of plaquettes not including $U_p$. Here $\chi_r(U)$ denotes
the character corresponding to the representation $r$. For SU(2), $r =
0,1/2,3/2,\ldots$ and for SU(3), $r= (\lambda,\mu)$ where $\lambda$
denotes the number of boxes in the first row of the Young tableau describing the representation and $\mu$ is the number of boxes in the second row. Similarly, for SU$(N)$, $r=(r_1,r_2,\ldots,r_{N-1})$.

Performing a character expansion of the exponential in \eqn{interest} gives:
\bea
\int dU_p e^{S(U_p)} \chi_{r}(U_p V) &=& \sum_{r'} \int dU_p
c_{r'} \chi_{r'}(U_p)\chi_{r}(U_p V).
\eea
This is simply a generalisation of a Fourier expansion.
Here, the coefficient $c_{r'}$ is given by:
\bea
c_{r'} = \int dU_p \chi_{r'}(U_p) e^{S(U_p)}.
\eea
Now, using the orthogonality relation,
\bea
\int dU_p
\chi_{r'}(U_p V) \chi_{r}(U_p) = \frac{1}{d_r} \delta_{r'r} \chi_r(V),
\label{charorthog}
\eea
where $d_r$ is the dimension of the representation $r$, we obtain:
\bea
\int dU_p e^{S(U_p)} \chi_{r}(U_p V) &=& 
\frac{1}{d_r} \chi_r(V)\int dU_p \chi_{r}(U_p) e^{S(U_p)}.
\label{integrate}
\eea
This result allows us to integrate out a single plaquette from an
extended Wilson loop in a given representation $r$. To complete the
calculation we need to relate SU($N$) characters to the 
traces of group
elements. This can be done using Weyl's character
formula~\cite{Bars:1980yy}. For SU($N$), according to Bars~\cite{Bars:1980yy}, the dimensions and characters corresponding to the first few representations are given by:
\bea
\begin{array}{rclrcl}
\displaystyle \chi_{1}(U) &=& \Tr U &
\displaystyle d_{1}(U) &=& N \vspace{0.1cm}\\
 \chi_{2}(U) &=& \displaystyle \frac{1}{2}\left[(\Tr U)^2 + \Tr U^2\right] &
d_{2}(U) &=& \displaystyle \frac{1}{2}N(N+1)\vspace{0.1cm}\\
\displaystyle\chi_{1 1}(U) &=&\displaystyle \frac{1}{2}\left[(\Tr U)^2 - \Tr U^2\right] &
d_{11}(U) &=&\displaystyle \frac{1}{2}N(N-1)\vspace{0.1cm} \\
\chi_{21}(U) &=&\displaystyle \frac{1}{3}\left[(\Tr U)^3 - \Tr U^3\right] &
d_{21}(U) &=&\displaystyle \frac{1}{3}(N-1)N(N+1)\vspace{0.1cm} \\
\displaystyle \chi_{1^{N-1}}(U) &=& \displaystyle \Tr U^\dagger &
\displaystyle d_{1^{N-1}}(U) &=& N
\end{array}
\label{sunchars}
\eea
Here we have adopted the convention of dropping all zeros in the
character labels.  
The Mandelstam constraints for the gauge group in question allows all 
characters to be expressed in terms of a minimal set of trace variables. 
For
example, for SU(3) we make use of the Mandelstam constraint,
\bea
\Tr (A^2 B) = \Tr A \Tr (A B) - \Tr A^\dagger \Tr B + \Tr (A^\dagger B),
\label{su3mand}
\eea  
where $A\in {\rm SU}(3)$ and $B$ is any $3\times 3$ matrix, to express all
characters in terms of $\Tr U$ and $\Tr U^\dagger$. For example,
for the case of SU(3), \eqn{sunchars} simplifies to  
\bea
\begin{array}{rclcrcl}
 \displaystyle \chi_{1}(U) &=& \Tr U &\qquad& d_{1} &=& 3\\ 
\vspace{0.1cm}
\displaystyle \chi_{11}(U) &=& \displaystyle
\displaystyle \frac{1}{2}\left[(\Tr U)^2-\Tr(U^2)\right]= \Tr U^\dagger
&\qquad & d_{11} &=& 3\\ 
\displaystyle \vspace{0.1cm}
\chi_{2}(U) &=& \displaystyle
\displaystyle\frac{1}{2}\left[(\Tr U)^2+\Tr(U^2)\right]= (\Tr U)^2-
\Tr U^\dagger  &\qquad & d_{2} &=& 6\\
\displaystyle  \chi_{21}(U) &=& \displaystyle
\displaystyle \frac{1}{3}\left[(\Tr U)^3-\Tr(U^3)\right]= \Tr U \Tr U^\dagger -1 & \qquad & d_{21} &=& 8
\end{array}
\label{su3chars}
\eea     
However, for general SU($N$) the Mandelstam constraints are
difficult to calculate.
In what follows we will need expressions for $\Tr U$, $\Tr U^\dagger$,
$\Tr (U^2)$, $(\Tr U)^2$, and $\Tr U \Tr U^\dagger$ as linear
combinations of characters for SU($N$). 
It is possible to do this without the use
of the Mandelstam constraint. Such expressions are necessary in order 
to make
use of \eqn{integrate} in
the calculation of expectation values of trace variables. For $\Tr U$,
$\Tr U^\dagger$, $\Tr (U^2)$ and $(\Tr U)^2$ the necessary expressions 
are easily obtained by rearranging
\eqn{sunchars},
\bea
\Tr U &=& \chi_{1}(U) \nn\\
(\Tr U)^2 &=&\chi_{2}(U)+ \chi_{11}(U) \nn\\
\Tr U^2 &=&\chi_{2}(U)-\chi_{11}(U) \nn\\
\Tr U^\dagger &=& \chi_{1^{N-1}}(U).
\label{tracevars-1}
\eea 
To express the remaining expression,
$\Tr U \Tr U^\dagger$, as a linear combination of characters is not as
easily done. For SU(3) one can simply rearrange
the expression for $\chi_{21}(U)$ in \eqn{su3chars}. 
For the general $N$ case 
it is simplest to consider Young tableaux. In terms of Young tableaux we 
have
\bea
\Tr U \Tr U^\dagger  \equiv \square \,\otimes \left.\begin{array}{c} \square\vspace{-0.18cm}\\
\square\vspace{-0.18cm} \\ \vdots \\
\square \end{array}\right\}N-1 .
\eea 
Performing the product representation decomposition gives
\bea
\square \,\otimes \left.\begin{array}{c} \square\vspace{-0.18cm}\\
\square\vspace{-0.18cm} \\ \vdots \\
\square \end{array}\right\}N-1 &=& \left.
\begin{array}{c} \square\vspace{-0.18cm}\\\square\vspace{-0.18cm}\\\vdots\\\square \end{array}\right\}N +
\left.\begin{array}{c c} \square\vspace{-0.18cm} &
\hspace{-0.27cm}\square\\\square\vspace{-0.18cm}\\\vdots\\\square \end{array}
\right\}N-1
\eea
Converting back into the notation of characters and traces gives
\bea
\Tr U \Tr U^\dagger &=&  1 + \chi_{2 1^{N-2}}(U).
\label{uudag}
\eea

Eqn(\ref{uudag}) together with \eqns{tracevars-1}{integrate}
allow the analytic calculation of each contribution to $D_{ll'}$ for
all $N$. For the case of $\Delta D_{l l'}$ described earlier, we have 
\bea
\Delta D_{l l'} = \frac{2}{N} F_{Z_1}(N_l+ N_{l'} - 2 N_{l\cap l'})
\left[F_{Z_1^2}(N_{l\cap l'}) +F_{Z_1\!\bar{Z}_1}(N_{l\cap l'}) \right] 
-4 F_{Z_1}(N_l)F_{Z_1}(N_{l'}),
\eea
where the character integrals are defined by:
\bea
F_{Z_1}(n) &\!\!\!=\!\!\!& 
\left(\frac{1}{N}\right)^{n-1} 
\langle Z_1 \rangle^n ,  \\
F_{Z_1^2}(n) &\!\!\!=\!\!\!& 
 \frac{1}{2}\left[\frac{1}{N(N+1)}\right]^{n-1}\langle Z_1^2
+ Z_2 \rangle^n  +
 \frac{1}{2}\left[\frac{1}{N(N-1)}\right]^{n-1}\langle Z_1^2
- Z_2 \rangle^n , \\
F_{Z_2}(n) &\!\!\!=\!\!\!& 
 \frac{1}{2}\left[\frac{1}{N(N+1)}\right]^{n-1}\langle Z_1^2
+ Z_2 \rangle^n  -
 \frac{1}{2}\left[\frac{1}{N(N-1)}\right]^{n-1}\langle Z_1^2
- Z_2 \rangle^n  , \\
F_{Z_1\bar{Z}_1}(n) &\!\!\!=\!\!\!& 
1 +
\left[\frac{1}{(N-1)(N+1)}\right]^{n-1} \left(\langle Z_1 \bar{Z}_1
\rangle - 1 \right)^n.
\label{charfunctions}
\eea
Here we have made use of the notation, $Z_n := \Tr\left( U^n\right)$, 
to denote the trace
variables occupying a single plaquette, $U$.
The expectation values appearing in \eqn{charfunctions} are easily
expressed in terms of the generating functions of
\eqns{coolsum}{coolersum}. Differentiating the generating functions
appropriately gives
\bea
\langle Z_1 \rangle &=&  \frac{1}{G_{{\rm SU}(N)}} \frac{\partial
G_{{\rm SU}(N)}}{\partial c}\Bigg|_{d=c} \nn\\
\langle Z_1 \bar{Z}_1\rangle &=& \frac{1}{G_{{\rm SU}(N)}} \frac{\partial^2
G_{{\rm SU}(N)}}{\partial c \partial d}\Bigg|_{d=c} \nn\\
\langle Z_1^2 \pm  Z_2\rangle &=& \frac{1}{G_{{\rm SU}(N)}} \frac{\partial^2
G_{{\rm SU}(N)}}{\partial c^2}\Bigg|_{d=c} \pm \frac{1}{H_{2}}  \frac{\partial
H_{2}}{\partial d}\Bigg|_{d=0}. 
\eea
In practice, we do not need to calculate all of
these matrix elements. 
We see from \eqn{coolsum} that $\langle Z_1^2\rangle$ and
$\langle Z_1 \bar{Z}_1 \rangle$ are related by 
\bea
\langle Z_1^2 \rangle &=& \frac{1}{2}\frac{d^2\langle Z_1
\rangle}{dc^2}-\langle Z_1 \bar{Z}_1 \rangle . 
\eea 
This follows from the fact that a group integral does not depend on
the choice of direction for the links. To be more precise, the result,
\bea
\int_{{\rm SU}(N)} d U f(U) =  \int_{{\rm SU}(N)} d U^\dagger
f(U^\dagger) = \int_{{\rm SU}(N)} d U f(U^\dagger),
\eea
follows from the fact that $dU^\dagger$ and $d U$ each define invariant Haar
measures on SU($N$) which, by uniqueness, must be equal.
 
We now move on to the calculation of $N_{ll'}$. It is easy to show
that the only non-zero contributions occur when there is at least one
common link and an overlap between the rectangles. The improvement
term (the second term in \eqn{N}) only contributes when the two
rectangles share at least two neighbouring links in a given
direction. Consider the contribution $\Delta N_{l l'}$ to $N_{ll'}$ in
which there are $L_1$ common links and $L_2$ common strings of two
links in a given direction. Again we suppose $N_l$ plaquettes are
enclosed by rectangle $l$ and that there are $N_{l\cap l'}$ common plaquettes. 
Making use of \eqn{integrate} and following \rcite{Carlsson:2002ss} we obtain
\bea
\Delta N_{ll'} = \frac{L}{N} F_{Z_1}(N_l\!+\!N_{l'}\!-\!2N_{l\cap l'})\left[ 
F_{Z_2}(N_{l\cap l'})
\!-\!\frac{1}{N} F_{Z_1^2}(N_{l\cap l'})
\!-\! N \!+\! \frac{1}{N} F_{Z_1\!\bar{Z}_1}(N_{l\cap l'})\right],
\eea
with
\bea
L = (1-\kappa)L_1 + \frac{\kappa}{u_0^2}L_2.
\eea
For the case of SU(3) we can simplify this using \eqn{su3mand} to 
\bea
\Delta N_{ll'} = \frac{L}{3} F_{Z_1}(N_l\!+\!N_{l'}\!-\!2N_{l\cap l'})\left[ 
\frac{2}{3}F_{Z_1^2}(N_{l\cap l'})
\!-\!2 F_{Z_1}(N_{l\cap l'})
\!-\! 3 \!+\! \frac{1}{3} F_{Z_1\!\bar{Z}_1}(N_{l\cap l'})\right].
\eea

Having determined individual contributions to $D_{l l'}$ and $N_{l
l'}$, to complete their calculation the possible
overlaps between states $l$ and $l'$ of a given type must be counted.

\subsection{Choosing an appropriate vacuum state}
\label{choosinganappropriate}
In \sect{fixingthevariational} we calculated variational vacuum
wave functions for pure SU($N$) gauge theory for $N=2$, 3, 4 and 5. Our
motivation was to use these wave functions as inputs to calculations
of SU($N$) massgaps. We obtained wave functions with a variational
parameter that was proportional to $\beta$ in the large $\beta$ limit
and $\beta^2$ in the small $\beta$ limit. However, this behaviour is
incompatible with the exact continuum vacuum wave function. For a
one-plaquette trial state, given by \eqn{oneplaquette}, to be compatible
with the exact continuum vacuum wave function  in 2+1 dimensions, one must have
$c\propto \beta^2$ in the large $\beta$ limit~\cite{Greensite:1987rg}.
This result is
independent of the dimension of the gauge group in question. For the
case of SU(2) in 2+1 dimensions, in the scaling
region it has been shown~\cite{Greensite:1987rg} that for compatibility with the exact SU(2) 
continuum vacuum wave function, we must use the Greensite vacuum wave function 
\be
\begin{array}{c}\includegraphics{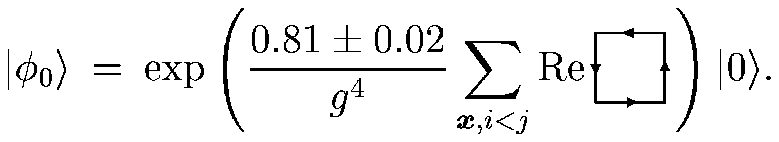}\end{array}
\label{greensite}
\ee
It is this vacuum wave function that was used by Arisue~\cite{Arisue:1990wv}
in the calculation 
that we generalise in this paper. 
It would thus seem that
using a variational wave function is not appropriate in the calculation
of massgaps. However, the scaling argument presented in
\rcite{Carlsson:2002ss} demonstrates that in a glueball mass scaling region the
precise form of $c(\beta)$ is not important. For this
reason we expect that the results presented here will agree with a
calculation using the Greensite wave function. 

As it turns out, to calculate variational wave functions 
for large dimension gauge groups is cumbersome. Numerical precision 
becomes a factor in the minimisation of the energy density. This 
problem is magnified in the calculation of tadpole improved results. We thus abandon the use
of variational wave functions in the calculation of massgaps beyond SU(5). Instead
we make use of the one plaquette wave function of \eqn{oneplaquette} and
define a simple dependence, $c(\beta)$, which most often will be
$c(\beta) =\beta$. Calculations for $N>5$ will be presented in a later 
publication.

\section{SU(2), SU(3), SU(4) and SU(5) Massgap Results}

In this section we present glueball mass results for SU($N$) pure
gauge theory in 2+1 dimensions with $N=2$, 3, 4 and 5.
For each SU(3) calculation we keep 80 terms in the $k$-sum of
\eqn{Y} giving convergence up to $\beta = 50$.  For $N>3$ the
truncation $l_{{\rm max}} = 20$ is used. The generation of 
$N_{ll'}$ and $D_{ll'}$  and 
implementation of the minimisation process is accomplished 
with a Mathematica code. 

\begin{figure*}
\centering
\subfigure[SU(2)] 
                     {
                         \label{su2conv}
                         \includegraphics[width=7cm]{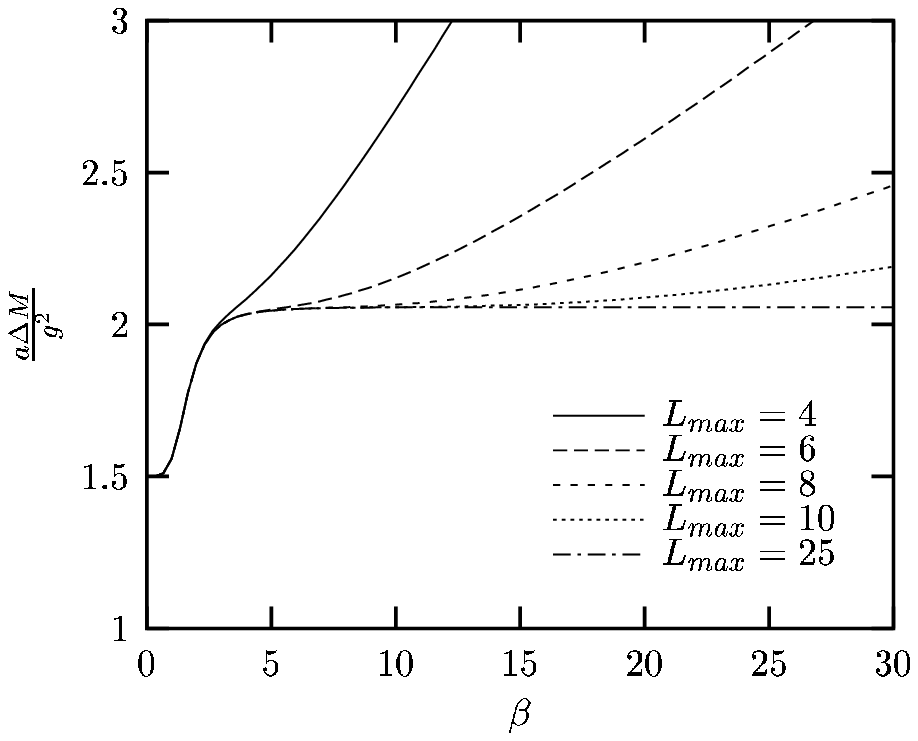}
                     }                   
 \subfigure[SU(3)] 
                     {
                         \label{su3conv}
                         \includegraphics[width=7cm]{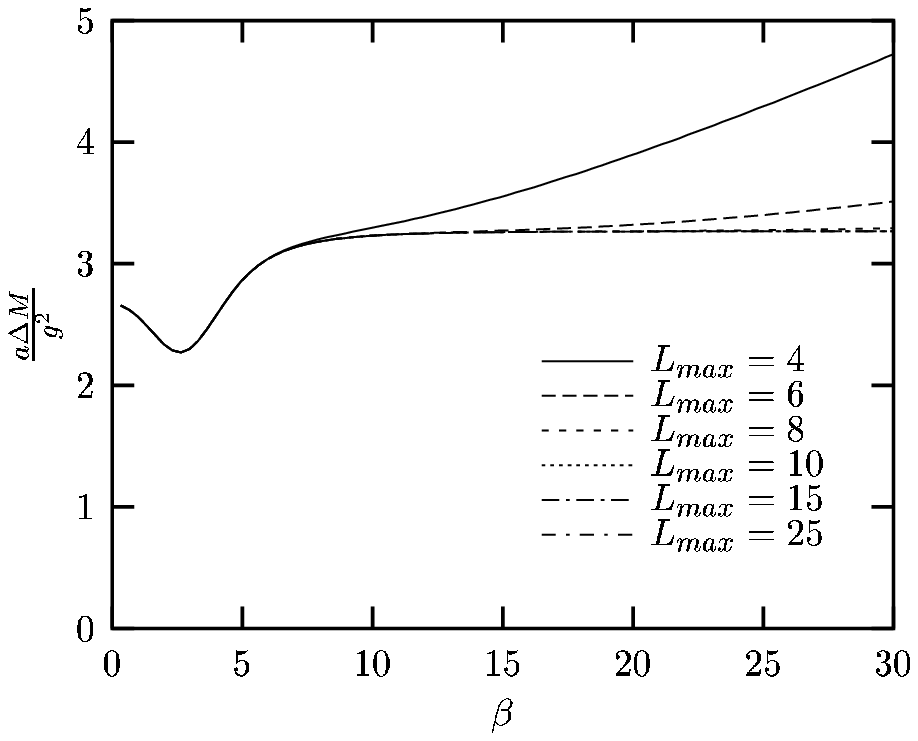}
                     }\\
 \subfigure[SU(4)] 
                     {
                         \label{su4conv}
                         \includegraphics[width=7cm]{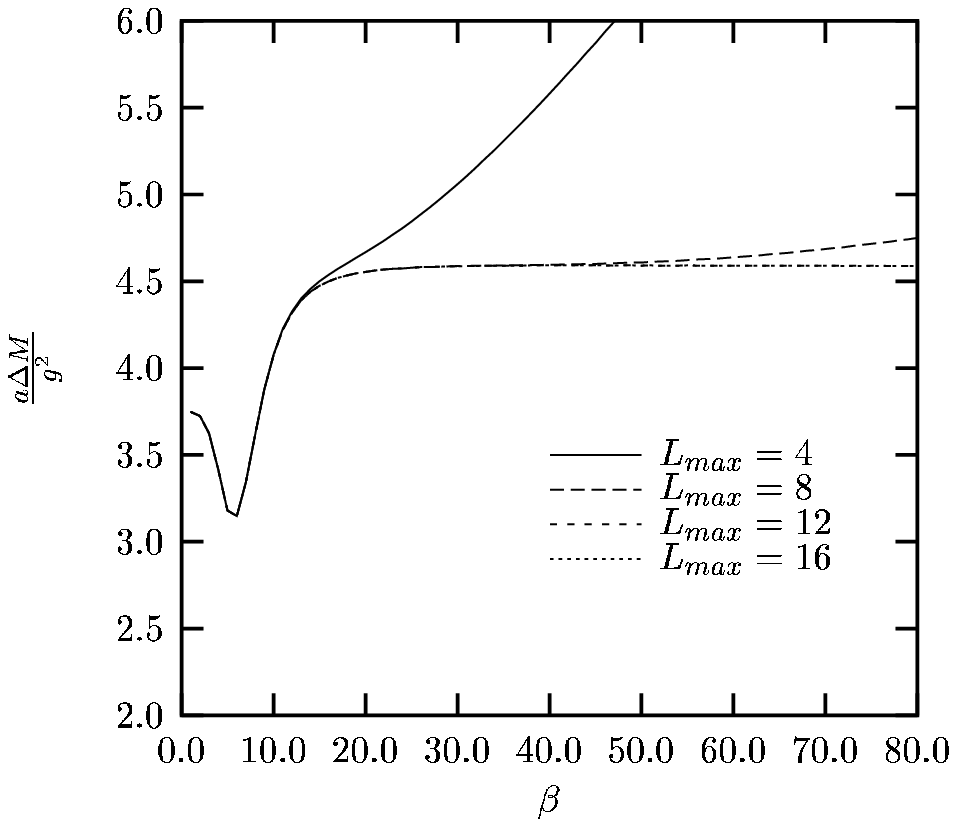}
                     }
\subfigure[SU(5)] 
                     {
                         \label{su5conv}
                         \includegraphics[width=7cm]{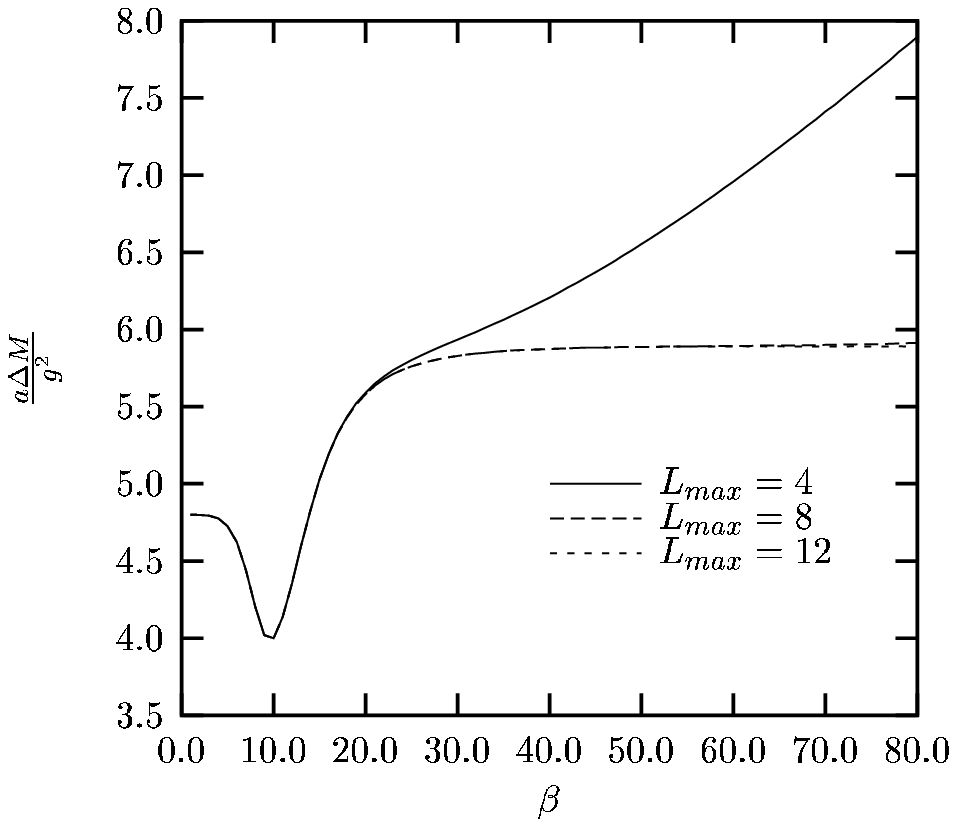}
                     }\caption{The unimproved 2+1 dimensional 
                     symmetric massgaps for SU(2), SU(3), SU(4) and SU(5).}

\label{mgconv}
\end{figure*}

For the case of 2+1 dimensions we expect $a \Delta m/g^2$ to become constant in the scaling region. 
The convergence of the massgaps with $L_{max}$ is illustrated in
\fig{mgconv}. We notice that for $N>2$ only small improvements to the
scaling behaviour are gained by extending the calculation beyond order
8 on the range of couplings shown. This suggests that a more complicated basis (including, for
example, nonrectangular loops) is required to simulate SU($N$)
excited states with $N>2$ than for the case of SU(2).

\begin{figure*}
\centering
\subfigure[Symmetric SU(2) massgap] 
                     {
                         \label{su2sym}
                        \includegraphics[width=7cm]{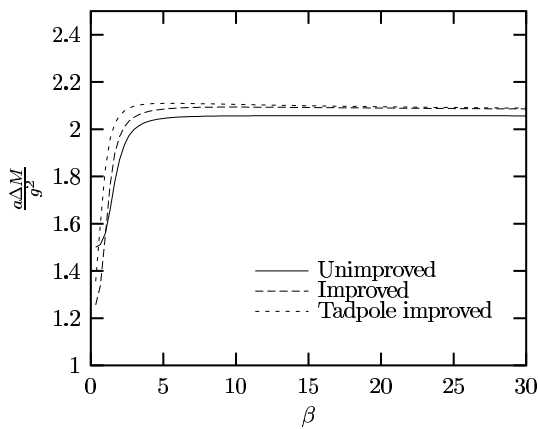}
                     }                   
 \subfigure[Symmetric SU(3) massgap] 
                     {
                         \label{su3sym}
        \includegraphics[width=7cm]{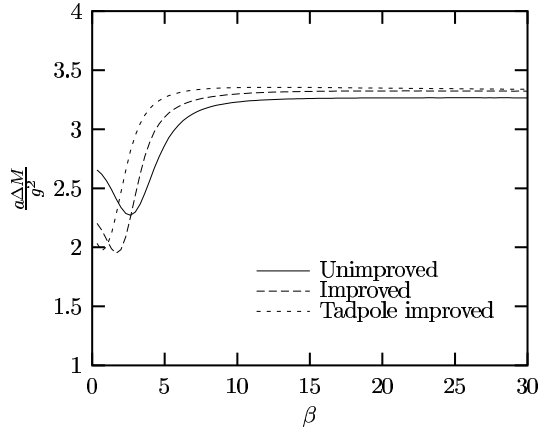}
                     }\\
\subfigure[Symmetric SU(4) massgap] 
                     {
                         \label{su4sym}
                        \includegraphics[width=7cm]{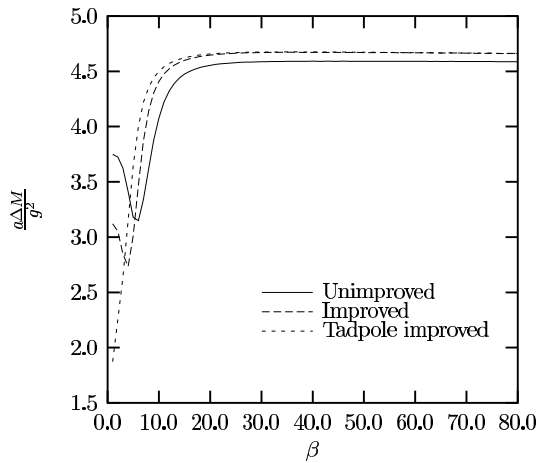}
                     }                   
 \subfigure[Symmetric SU(5) massgap] 
                     {
                         \label{su5sym}
                         \includegraphics[width=7cm]{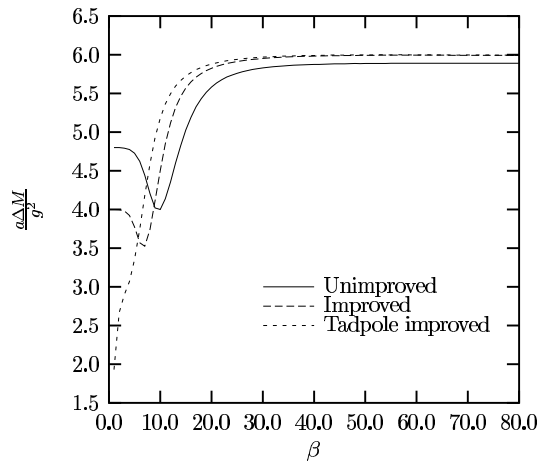}
                     }\\
\caption{2+1 dimensional symmetric massgaps for SU(2), SU(3) (both with
                         $L_{max}=25$), SU(4) (with $L_{max}=16$) and
                         SU(5) (with $L_{max} = 12$).}
\label{mgcompare}
\end{figure*}

\begin{figure*}
\centering
\subfigure[Antisymmetric SU(3) massgap] 
                     {
                         \label{su3asym}
                        \includegraphics[width=7cm]{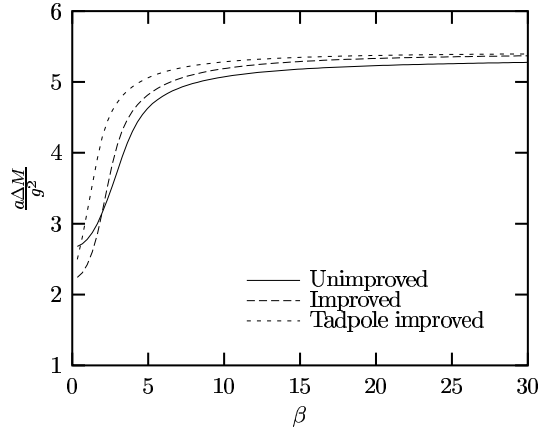}
                     } \\
\subfigure[Antisymmetric SU(4) massgap] 
                     {
                         \label{su4asym}
                        \includegraphics[width=7cm]{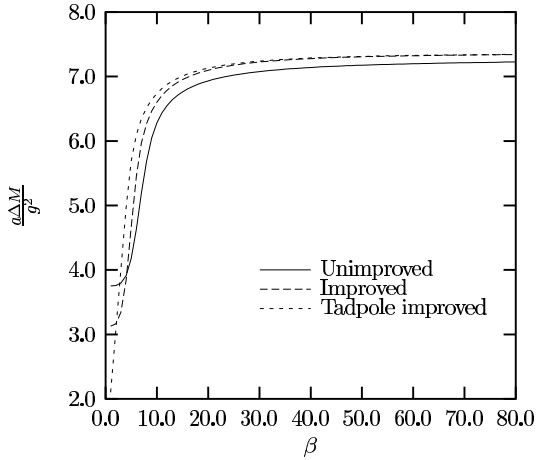}
                     } 
 \subfigure[Antisymmetric SU(5) massgap] 
                     {
                         \label{su5antisym}
                         \includegraphics[width=7cm]{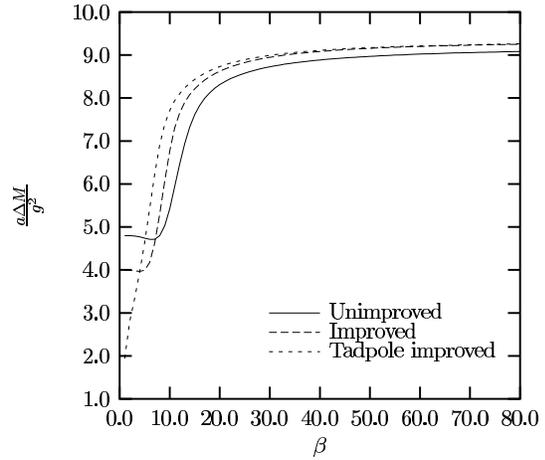}
                     }
\caption{2+1 dimensional antisymmetric massgaps for SU(3) (with
                         $L_{max}=25$), SU(4) and SU(5) (both with $L_{max}=12$).}
\label{asymmgcompare}
\end{figure*}

In \fig{mgcompare} results for the lowest lying glueball mass, 
calculated with Kogut-Susskind, improved and
tadpole Hamiltonians, are shown.  
We see that $a \Delta m^S_1/g^2$ is approximated
well by a constant, in very large scaling regions, for the lowest lying
eigenstates for all $N$ considered. The scaling behavior becomes
significantly worse for the antisymmetric sector which is shown in
\fig{asymmgcompare} and for higher energy eigenvalues. This is because
the simplistic form of our excited state wave function is not
sufficient to reproduce the plaquette correlations required to
simulate these higher order states. One would expect the simulation of
higher order eigenstates to improve by including more complicated
loops in our expansion basis or by using a more complicated ground
state. The continuum limit excited states results for SU($N$) are given in \tabss{mgg}{mgmgmg}.

For the unimproved SU(2) case, the masses of the lowest two
eigenstates agree closely with 
the calculations of Arisue~\cite{Arisue:1990wv} (respectively $2.056\pm
0.001$ and $3.64\pm 0.03$ in units of $e^2=g^2/a$) in which the Greensite
vacuum wave function of \eqn{greensite} was used. This serves as a
check on our counting in calculating the possible overlaps of excited
states. Our calculation is in disagreement with that of Arisue at the
third eigenstate, for which Arisue calculates a mass $(5.15\pm
0.1)e^2$. Our fourth eigenstate is close in mass to Arisue's third and our
third eigenstate does not appear in his results. The reasons for this
are not clear. 

The results for the SU(3) symmetric massgap (in units of
$e^2$)
are to be compared to calculations by Luo and Chen $2.15 \pm 0.06$~\cite{Luo:1996ha}, Samuel
$1.84 \pm 0.46$~\cite{Samuel:1997bt} and Teper $2.40 \pm
0.02$~\cite{Teper:1998te}. Our result of $3.26520 \pm 0.00009$ is
considerably higher than all existing comparable results. By including
more complicated loops in the expansion basis one would expect to
reduce this estimate. This is emphasised by the fact that when using
only square basis states our result is considerably higher. To explain
the discrepancy between the our results and other's it is important to
note that since we use a basis of rectangles we exclude the
contribution of many non-rectangular small area diagrams that are
included in the calculations of Teper and that of Luo and Chen. For
this reason it may be the case that what we have interpretted as the
lowest glueball mass in this paper may, in fact, be a higher order
excited state. Teper has calculated the masses of the three lowest mass glueballs for SU(3) in
the $0^{++}$ sector~\cite{Teper:1998te}: $0^{++}$, $0^{++*}$ and $0^{++**}$,  with the respective results, in units of
$e^2$: $2.40\pm 0.02$, $3.606 \pm  0.063$ and $4.55 \pm
0.11$. It is interesting to note that our result is closer to Teper's
first excited state. In the same study Teper also calculated glueball masses in the $0^{++}$ sector for $N=4$, 5 and 6. The mass, in units of $e^2$, 
of his $0^{++*}$ state for SU(4) is $4.84 \pm 0.12$ and $5.99 \pm
0.16$ for SU(5). We notice that as $N$ is increased the results
presented here move closer to the mass of Teper's $0^{++*}$ state, with the improved
results being closer than the unimproved. In fact for SU(5), the
results presented  here, improved and unimproved, are consistent with Teper's $0^{++*}$ mass.
This forces us to question the interpretation of the large $\beta$
plateaux in \fig{mgcompare} as scaling regions for the 
lowest mass glueballs. It is
possible that the minima present in \fig{mgcompare} in the small
$\beta$ region are possible scaling regions. It is possible that 
our vacuum wave function and minimisation basis are insufficient to 
extend this scaling region over a wide range of couplings and that as
our approximation breaks down we observe a level crossing effect. 
We will 
examine the possibility of the small $\beta$ minima being scaling
regions in a later publication.

The antisymmetric results presented here can also be compared with
those of Teper~\cite{Teper:1998te}. While each of our results is
considerably higher than the masses of Teper's $0^{--}$ and $0^{--*}$
states, Teper's $0^{--**}$ state has a mass which is close to the lowest
mass state calculated here. Teper obtains the following masses, in
units of $e^2$, for the $0^{--**}$ state: $5.42\pm 0.16$ for
SU(3), $6.98 \pm 0.26$ for SU(4) and $9.18\pm 0.45$ for
SU(5). It is interesting to note that our corresponding 
lowest unimproved glueball
masses are consistent with these results. Our improved results
show better agreement with Teper's $0^{--**}$ state for SU(3) and
SU(5) than the corresponding unimproved results. The improved SU(4)
results are also consistent with Teper's SU(4) $0^{--**}$ mass
although the agreement is closer for the unimproved result.

\begin{table}
\caption{Estimates of the lowest lying SU(2) glueball masses (in units
of $e^2$) computed with various Hamiltonians in 2+1 dimensions. The
unimproved, improved and tadpole results are calculated in the
respective scaling regions $13.5 \le \beta \le 30.0$, $9.9 \le \beta
\le 30.0$ and $9.25 \le \beta \le 30.0$.} 
\label{mgg}
\begin{ruledtabular}
\begin{tabular}{cD{/}{\mbox{$\pm$}}{-2}D{/}{\mbox{$\pm$}}{-2}D{/}{\mbox{$\pm$}}{-2}}
 & \multicolumn{1}{c}{Unimproved} & \multicolumn{1}{c}{Improved} & \multicolumn{1}{c}{Tadpole Improved}   \\
\hline
$\Delta m^S_1$ & 2.05691/0.00002 &  2.0897/0.0003 
& 2.0965/0.0006 \\
$\Delta m^S_2$ & 3.645/0.001 &  3.685/0.001 
& 3.6953/0.0009 \\
$\Delta m^S_3$ & 4.5202/0.0004 & 4.574/0.004 
& 4.583/0.004 \\
$\Delta m^S_4$ & 5.133/0.003  & 5.177/0.004 
& 5.189/0.004  \\
$\Delta m^S_5$ & 5.867/0.006 & 5.932/0.008 
& 5.943/0.008 \\
\end{tabular}
\end{ruledtabular}
\end{table} 

\begin{table}
\caption{Estimates of the lowest lying symmetric SU(3) glueball masses (in
 units of $e^2$) computed with various Hamiltonians in 2+1
 dimensions. The results are calculated in the scaling region which
 minimises the standard error in each case.}\label{mgmg} 
\begin{ruledtabular}
\begin{tabular}{cD{/}{\mbox{$\pm$}}{-2}D{/}{\mbox{$\pm$}}{-2}D{/}{\mbox{$\pm$}}{-2}}
 & \multicolumn{1}{c}{Unimproved} & \multicolumn{1}{c}{Improved} & \multicolumn{1}{c}{Tadpole Improved}   \\
\hline
$\Delta m^S_1$ & 
3.265868  / 0.000042 &
3.32365  /  0.00012 &
3.32580  /  0.00015 \\
$\Delta m^S_2$ &
6.23903  /  0.00065 &
6.30391  /  0.00083 &
6.31192  /  0.00084 \\
$\Delta m^S_3$ & 
7.5767  /  0.0025 &
7.6466  /  0.0030 &
7.6498 /  0.0030  \\
$\Delta m^S_4$ & 
8.9462  /  0.0029  &
9.0118  /  0.0044  &
9.0206  /  0.0045 \\
$\Delta m^S_5$ &
10.0778 /  0.0071  &
10.1546  /  0.0094  &
10.1628  /  0.0094  \\
\end{tabular}
\end{ruledtabular}
\end{table}

\begin{table}
\caption{Estimates of the lowest lying symmetric SU(4) massgaps (in
units of $e^2$) computed with various Hamiltonians in 2+1
dimensions. The results are calculated in the scaling region which
minimises the standard error in each case.} 
\begin{ruledtabular}
\begin{tabular}{cD{/}{\mbox{$\pm$}}{-2}D{/}{\mbox{$\pm$}}{-2}D{/}{\mbox{$\pm$}}{-2}}
 & \multicolumn{1}{c}{Unimproved} & \multicolumn{1}{c}{Improved} & \multicolumn{1}{c}{Tadpole Improved}   \\
\hline
$\Delta m^S_1$ &
4.59121/ 0.00007       & 4.6720/  0.0001    & 4.6754/  0.0001 \\
$\Delta m^S_2$ &
 8.8122/ 0.0012        & 8.9276/  0.0016     & 8.9284/  0.0017 \\
$\Delta m^S_3$ &
10.5889/ 0.0051         & 10.7807/  0.0051      & 10.7794/  0.0051\\
$\Delta m^S_4$ &
12.5527/ 0.0048        & 12.6266/  0.0081      & 12.6138/  0.0080 \\
$\Delta m^S_5$ &
 14.052/ 0.012   & 14.165/  0.016       & 14.157/  0.016 \\
\end{tabular}
\end{ruledtabular}
\end{table} 

\begin{table}
\caption{Estimates of the lowest lying symmetric SU(5) massgaps (in
units of $e^2$) computed with various Hamiltonians in 2+1
dimensions. The results are calculated in the scaling region which
minimises the standard error in each case.}
\begin{ruledtabular}
\begin{tabular}{cD{/}{\mbox{$\pm$}}{-2}D{/}{\mbox{$\pm$}}{-2}D{/}{\mbox{$\pm$}}{-2}}
 & \multicolumn{1}{c}{Unimproved} & \multicolumn{1}{c}{Improved} & \multicolumn{1}{c}{Tadpole Improved}   \\
\hline
$\Delta m^S_1$ &
 5.8903/  0.0001 &  5.99434/  0.00009&
5.9983/ 0.0002
  \\
$\Delta m^S_2$ &
11.2335/  0.0036 &    11.3696/  0.0050 &  11.3731 /0.0049  \\
$\Delta m^S_3$ &
13.340/  0.011&   13.658/  0.012 &
13.663  / 0.011  \\
$\Delta m^S_4$ &
15.881/  0.012 &   15.890/  0.019&  
15.890/ 0.019  \\
$\Delta m^S_5$ &
17.564/  0.025 &  17.676/  0.035 &  
17.682/ 0.035  \\
\end{tabular}
\end{ruledtabular}
\end{table}

\begin{table}
\caption{Estimates of the lowest lying antisymmetric SU(3) massgaps
(in units of $e^2$) computed with various Hamiltonians in 2+1
dimensions. The results are calculated in the scaling region which
minimises the standard error in each case.} 
\begin{ruledtabular}
\begin{tabular}{cD{/}{\mbox{$\pm$}}{-2}D{/}{\mbox{$\pm$}}{-2}D{/}{\mbox{$\pm$}}{-2}}
 & \multicolumn{1}{c}{Unimproved} & \multicolumn{1}{c}{Improved} & \multicolumn{1}{c}{Tadpole Improved}   \\
\hline
$\Delta m^A_1$ & 5.32750 / 0.00047 &5.39661 /  0.00034
&5.39864 /  0.00032  \\
$\Delta m^A_2$ & 7.9389 /  0.0021 &8.0142 /  0.0028 &8.0145
/ 0.0028  \\
$\Delta m^A_3$ & 8.9319 /  0.0045 &9.0092 /  0.0056 &9.0087 /  0.0055 \\
$\Delta m^A_4$ & 10.4711 / 0.0058 &10.5514 / 0.0085 &10.5502 /  0.0085 \\
$\Delta m^A_5$ & 11.304 / 0.011 &11.384 / 0.015 &11.381 /  0.015 \\
\end{tabular}
\end{ruledtabular}
\end{table}

\begin{table}
\caption{Estimates of the lowest lying antisymmetric SU(4) massgaps
(in units of $e^2$) computed with various Hamiltonians in 2+1
dimensions. The results are calculated in the scaling regions which
minimise the standard error in each case.} 
\begin{ruledtabular}
\begin{tabular}{cD{/}{\mbox{$\pm$}}{-2}D{/}{\mbox{$\pm$}}{-2}D{/}{\mbox{$\pm$}}{-2}}
 & \multicolumn{1}{c}{Unimproved} & \multicolumn{1}{c}{Improved} & \multicolumn{1}{c}{Tadpole Improved}   \\
\hline
$\Delta m^A_1$ &
7.21479/  0.0012  &7.3310/  0.0011 &7.33586/
0.00077 \\
$\Delta m^A_2$ &
10.9117/  0.0033 & 11.0099/  0.004909&11.0617/
0.0046  \\
$\Delta m^A_3$ &
12.121/  0.007 &12.2779/  0.0088&12.292/
0.009 \\
$\Delta m^A_4$ &
14.5012/  0.0089 &14.592/  0.014&14.574/
0.015 \\
$\Delta m^A_5$ &
15.4521/  0.0173 & 15.545/  0.023&15.555/
0.023 \\
\end{tabular}
\end{ruledtabular}
\end{table}

\begin{table}
\caption{Estimates of the lowest lying antisymmetric SU(5) massgaps
(in units of $e^2$) computed with various Hamiltonians in 2+1
dimensions. The results are calculated in the scaling regions which
minimise the standard error in each case.} \label{mgmgmg}
\begin{ruledtabular}
\begin{tabular}{cD{/}{\mbox{$\pm$}}{-2}D{/}{\mbox{$\pm$}}{-2}D{/}{\mbox{$\pm$}}{-2}}
 & \multicolumn{1}{c}{Unimproved} & \multicolumn{1}{c}{Improved} & \multicolumn{1}{c}{Tadpole Improved}   \\
\hline
$\Delta m^A_1$ &
 9.067/  0.003&  9.239/  0.002 & 9.248 / 0.002 \\
$\Delta m^A_2$ &
13.717/  0.008&  13.89/  0.01&  13.89 / 0.01 \\
$\Delta m^A_3$ &
15.054/  0.015 & 15.32/  0.02&  15.327 / 0.019 \\ 
$\Delta m^A_4$ &
18.08/  0.02& 18.12/  0.03&  18.12 / 0.03 \\
$\Delta m^A_5$ &
19.084/  0.036 & 19.19/  0.05 & 19.20 / 0.05 \\ 
\end{tabular}
\end{ruledtabular}
\end{table}

When compared to equivalent unimproved calculations, the improved and
tadpole improved massgaps approach scaling faster as $\beta$ is
increased. This is evident in \figs{mgconv}{mgcompare} and is expected since,
for an improved calculation one is closer to the continuum limit when
working at a given coupling. However, for most improved calculations
the scaling behaviour is marginally less precise than the equivalent unimproved
calculation. A possible reason for this is that the one plaquette trial state
used here does not allow for direct contributions from the improvement term in
the kinetic Hamiltonian. For this term to contribute directly one would need a
trial state which includes Wilson loops extending at least two links
in at least one direction. 

The improved SU(2) massgap can be compared to the coupled cluster 
calculation of Li et al~\cite{Li:2000bg}. Their result (in units of $e^2$),
$\Delta m^S_1 = 1.59$, is again significantly lower than our result
$2.0897\pm0.0003$. The difference is again attributable to the different
choices of Wilson loops used in the simulation of states. While our
calculation makes use of the simple one plaquette ground state and a
minimisation basis with only rectangular loops, the coupled cluster 
calculation of Li et al uses a more accurate ground state wave
function consisting of an exponential of a sum of extended loops which
are not necessarily rectangular. Without including additional small
area Wilson loops we cannot be confident that the lowest mass state
accessible with our minimisation basis is in  fact the lowest mass
state of the theory. Clearly there is scope for more work here.

\section{Conclusion}
\label{concl}
In this paper we have extended the analytic techniques of 2+1
dimensional Hamiltonian LGT, traditionally used for SU(2), to general SU($N$).  Impressive scaling is achieved over an extremely wide range
of couplings for the lowest energy eigenstates in the symmetric and
antisymmetric sectors.  Our calculations use a one plaquette trial state
and a basis of rectangular states over which excited state energies
are minimised.  Such choices allow the use of analytic techniques in
SU($N$) calculations. 

The results of this paper give estimates of
the lowest unimproved, improved and tadpole improved SU($N$) 
glueball masses, all of
which are above current estimates. We suspect that the reason for the
discrepancy is a lack of small area non-rectangular states in our
minimisation basis. A basis of rectangular states was used for
simplicity. The inclusion of non-rectangular states is
straightforward and only complicates the counting of overlaps between
diagrams of a particular type.  When not including sufficient small area
diagrams it is possibly that the lowest mass states of the theory are not
accessible over a large range of couplings. A further improvement to
our calculation would involve the use of an improved trial vacuum
state. Such a state would possibly include several 
extended Wilson loops in its exponent. Without the development of new
techniques for performing the required integrals the use of such a
vacuum wave function would require the use of Monte Carlo techniques
for the calculation of expectation values. In this scenario many
advantages of the Hamiltonian approach would be lost.

In a later publication we extend the calculations presented here to
SU(25) in an attempt to explore the mass spectrum in the large $N$
limit of pure SU($N$)  gauge theory.

\bibliography{thesis-paper1}

\end{document}